\documentclass[aps,prb,showpacs,twocolumn]{revtex4-1}
\usepackage[dvips]{graphicx}
\usepackage{amsmath}
\usepackage{amssymb}
\usepackage{amsfonts}
\usepackage{bm}
\usepackage{textcomp}
\usepackage{latexsym}
\usepackage{color}

\newcommand*{\PS}{{$\rm P_2S_6$}{~}}
\newcommand*{\SPS}{{$\rm Sn_2P_2S_6$}{~}}
\newcommand*{\SPSe}{{$\rm Sn_2P_2Se_6$}{~}}
\begin{document}

\title{Electronic Structure and Phase Transition in Ferroelectic $\rm \mathbf{Sn_2P_2S_6}$ Crystal}
\date{\today}
\author{Konstantin \surname{Glukhov} }
\email{kglukhov@gmail.com}
\author{Kristina \surname{Fedyo} }
\author{Yulian \surname{Vysochanskii} }
\email{vysochanskii@gmail.com} %
\affiliation{\rm Uzhgorod National University, Institute for Solid
State Physics and Chemistry, 54, Voloshyn Str.,  Uzhgorod 88000,
Ukraine}

%\title{Electronic Structure and Phase Transition in $\rm Sn_2P_2S_6$ Ferroelectic Crystal}
%%\titlerunning{Electronic structure and phase transition in $\rm Sn_2P_2S_6$}
%\author{%
%  Konstantin Glukhov\textsuperscript{\Ast},
%  Kristina Fedyo,
%  Yulian Vysochanskii}
%%\authorrunning{K. Glukhov et al.}
%%E-mail-address of corresponding author
%\mail{e-mail
%  \textsf{kglukhov@gmail.com}, Phone:
%  +xx-xx-xxxxxxx, Fax: +xx-xx-xxx}
%
%  \institute{%
%  \textsuperscript{1}\,Uzhgorod National University, Institute for Solid
%State Physics and Chemistry, 54, Voloshyn Str.,  Uzhgorod 88000,
%Ukraine}
%
%\received{XXXX, revised XXXX, accepted XXXX} % do not change, will be filled in by the publisher
%\published{XXXX} % do not change, will be filled in by the publisher

% Please select about four verbal keywords for your manuscript.
%\keywords{\emph{Ab initio} calculations, phase transitions,
%electronic energy structure.}

%\abstract{%
% This is a macro for the typesetting of two-column text in an
% abstract. It will typeset the two arguments in \abstcol{}{} as the
% left and right column inside the abstract box. At the
% columnbreak there will be always a columnbreak (\par), so both
% columns start with a new paragraph. No automatic column height
% balancing is done.
%
% If used with a \titlefigure it will silently output both
% parameters as consecutive paragraphs.
%
% The macro is defined exclusively inside the argument of \abstract{};
% if used outside it will raise an error.
%
% Usage: \abstcol{<left column>}{<right column>}
\begin{abstract}
An analysis of the \PS cluster electronic structure and its
comparison with the crystal valence band in the paraelectric and
ferroelectric phases has been done by first-principles calculations
for \SPS ferroelectrics. The origin of ferroelectricity has been
outlined. It was established that the spontaneous polarization
follows from the stereochemical activity of the electron lone pair
of tin cations what is determined by hybridization with \PS
molecular orbitals. The chemical bonds covalence increase and
rearrangement are related to the valence band changes at transition
from the paraelectric phase to the ferroelectric one.
\end{abstract}


\maketitle

\section*{INTRODUCTION}

For such perovskite ferroelectrics as $\rm BaTiO_3$, the main origin
of spontaneous polarization is commonly related to the hybridization
interaction between the transition-metal and oxygen
ions~\cite{Cohen92}. Another mechanism involves cations with ''lone
pair'' electrons which have a formal $ns^2$ valence electron
configuration~\cite{Seshadri01}. In the same manner as for the
$d^{0}$ transition-metal ions, these $p^0$ ions (as example $\rm
Pb^{2+}$ for $\rm PbTiO_3$, or $\rm Bi^{3+}$ for $\rm BiMnO_3$)
contain some $p$-charge density which contribute to the displacive
distortions. If the lowering of energy associated with the
hybridization interaction is larger than the interionic repulsion
opposing the ion shift, then a ferroelectric distortion appears.
This ''stereochemical activity of the lone pair'' is the driving
force for off-center distortion in ferroelectrics. Both named
origins of ferroelectricity (the first~-- $d^0$-''ness'', and the
second~-- ''lone pair'' activity) are familiar to the second-order
Jahn-Teller (SOJT) effect~\cite{Bersuker01,Rondinelli09}. This
effect is determined by a balance of positive and negative
contributions to the total energy. The first one describes short
range repulsive forces and is related to the rigid ions (with frozen
electronic configuration) shifts from original high symmetry
positions. Such term is small for the cases of ''closed-shell''
$d^0$ or $p^0$ cations. The second, negative, contribution describes
the relaxation of electronic configuration in response to the ions
displacements through covalent bonds formation. This term favors the
ferroelectric distortion. For full picture, the geometrical (or
hybrid improper) mechanism, which is related to the rotational modes
that trigger instability of polar
mode~\cite{Rondinelli11,Fukushima11}, could also be considered at
study of ferroelectricity nature.

\begin{figure}[h!]
\hspace{-0cm}\center
\includegraphics[width=.90\columnwidth]{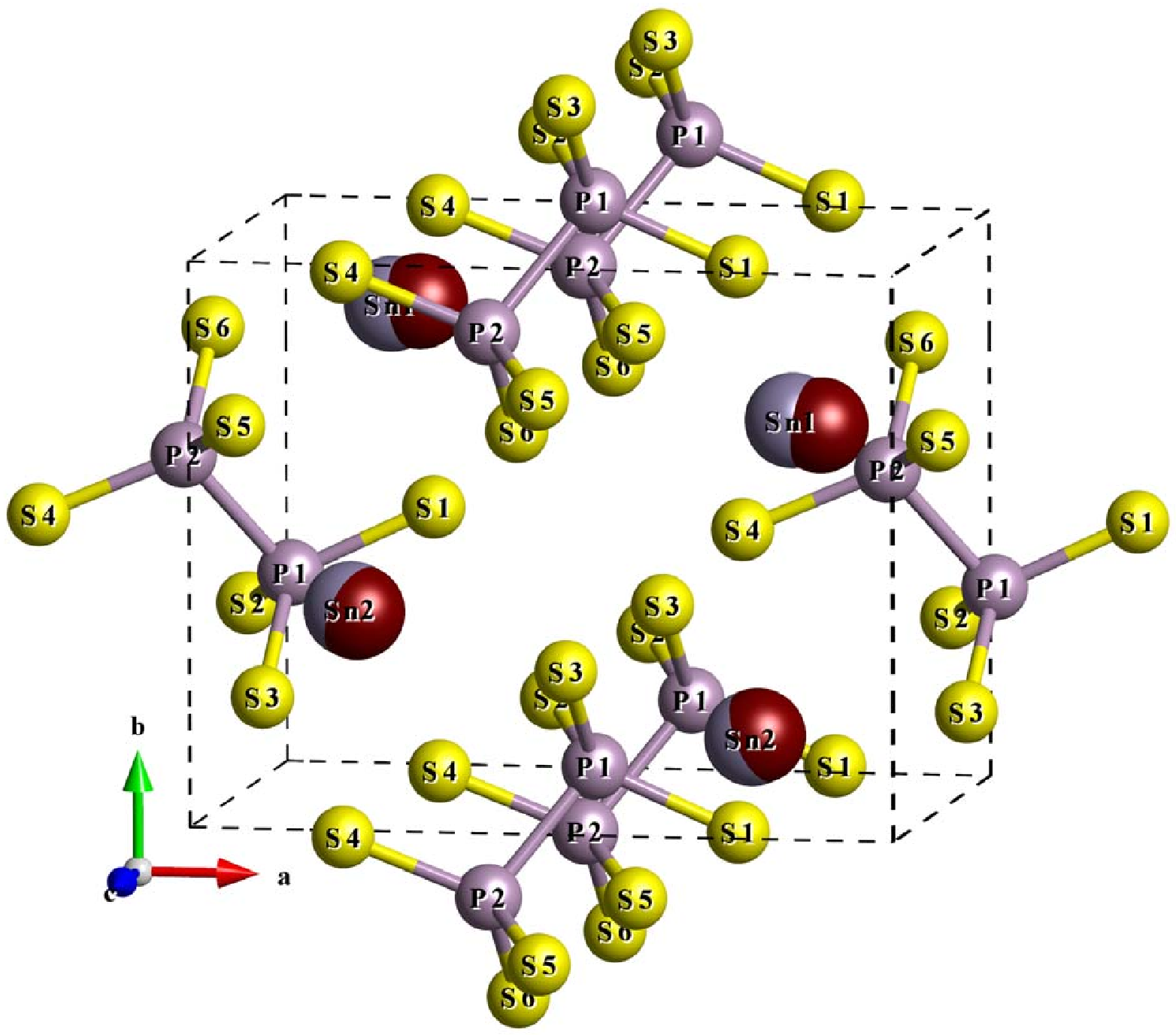}
\hfil\vspace{0.5cm}
\includegraphics[width=.45\columnwidth]{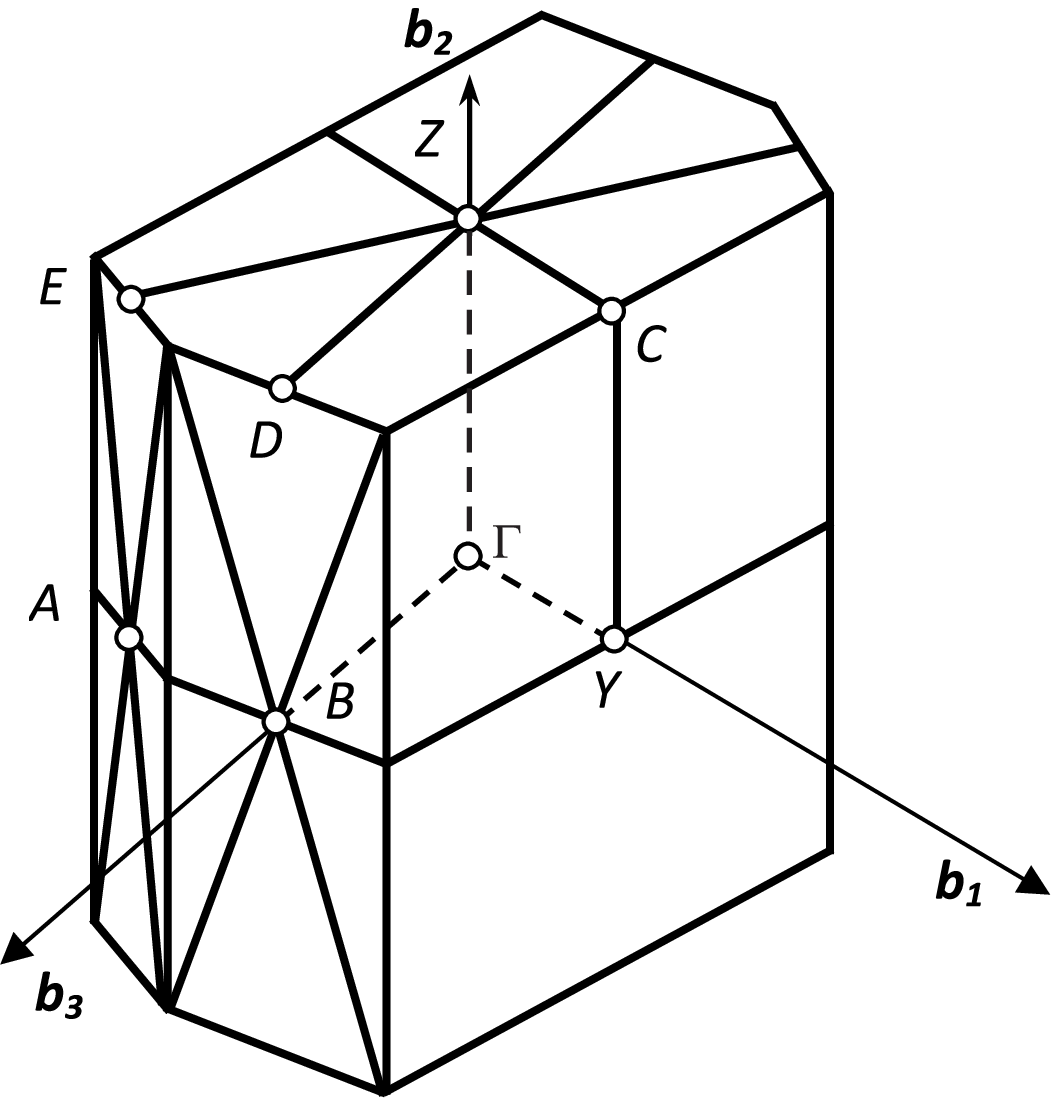}
\caption{\label{fig1}The crystal structure of \SPS ferroelectric phase~\cite{Dittmar74}. The tin atoms positions in the paraelectric
phase~\cite{Cleary92} are shown by red. The shape of
the Brillouin zone with denoted symmetrical points is shown for
primitive monoclinic lattice.}
\end{figure}

The cubic crystal lattices of $\rm ABO_3$ compounds are built by
covalent bonds A~---~O and B~---~O with considerable contribution of
ionicity for former ones. Naturally more complex bounding evolution
could be supposed at the ferroelectric phase transition in
ion-covalent crystal \SPS with monoclinic lattice. For this compound
the $\rm Sn^{2+}$ cations and the $\rm (P_2S_6)^{4-}$ anion clusters
are joined by mostly ionic Sn~---~S bonds at covalent P~---~S and
P~---~P bonding. \SPS uniaxial ferroelectric undergoes the second
order phase transition at $T_0\approx337$~K ($P2_1/c \rightarrow
Pc$, two formula units in the elementary cell for both phases
(Fig.~\ref{fig1}%~\cite{Dittmar74,Cleary92}
)) in a crossover displacive-order/disorder region~\cite{Vysochanskii06}.
Ferroelectric instability in \SPS crystal is a result of non-linear
$A_gB_u^2$ coupling of the soft polar $B_u$ and fully symmetrical
$A_g$ optic modes, leading to three-well
potential~\cite{Rushchanskii07}. Here the opposite picture appears
in compare with perovskites where the only one lattice mode could
determine dynamical instability related to the ferroelectric phase
transition~\cite{Rushchanskii07}. In general, all the 13 $B_u$ and
15 $A_g$ optic modes were accounted in the frozen phonons
approximation for construction of effective Hamiltonian for \SPS
crystal what was applied for the MC simulation of the ferroelectric
phase transition and their behavior under hydrostatic
pressure~\cite{Rushchanskii11p}. A system with a three-well
potential was early considered by Lines~\cite{Lines69,Lines70} with
application for the $\rm LiNbO_3$ and $\rm LiTaO_3$ crystals. Such
system can be generally described by two order parameters (related
to dipole and quadruple moments), and as a result, a variety of
stable, metastable and unstable states can be realized on a phase
diagrams~\cite{Hoston91,Ekiz01}.

The strong anharmonicity of \SPS crystal lattice is obviously  joined
with effective electron-phonon interaction, that appears as
a stereochemical activity of the tin cations electron lone pair
$5s^2$, and in fact, it is a reflection of the SOJT effect. Possible
leading role of the cations'{~} stereoactivity for \SPS crystals
was early noted at X-ray structure investigations of their
paraelectric phase in comparison with structure data for the
ferroelectric phase~\cite{Cleary92}. Structural evidences of the tin
cation stereoactivity was also analyzed in details by structure
refinement of the paraelectric and the ferroelectric phases for $\rm Sn_2P_2Se_6$ selenide
analog~\cite{Enjalbert99}. By M\"{o}ssbauer effect
investigations for $\rm {}^{119}Sn$ nucleus~\cite{Vysochanskii09}
and by NMR spectroscopy for isotopes $\rm{}^{31}P$ and $\rm
{}^{119}Sn$~\cite{Bourdon02,Apperley93}, the important changes of
chemical bonding at the ferroelectric phase transition in \SPS were
found. The X-ray photoelectron spectroscopy confirms growth of the
chemical bonds covalence in the ferroelectric phase~\cite{Grigas09}.

\SPS crystals are ferroelectric-semiconductors with promising
photorefractive~\cite{Grabar06}, photovoltaic~\cite{Cho01},
electrooptic~\cite{Haertle03} and piezoelectric~\cite{Maior00}
characteristics. Their ferroelectric properties are effectively
influenced by state of electronic subsystem~\cite{Molnar95}.
Influence of the sulfur and tin vacancies on semiconductive and
optic properties of \SPS crystals was studied
recently~\cite{Vysochanskii11}. These data motivate the electronic
structure investigation for \SPS crystals in the paraelectric and the
ferroelectric phases.

The first-principles calculations in LDA approach of Density
Functional Theory (DFT) for \SPS ferroelectric phase were carried
out by several groups~\cite{Lavrentyev03,Curro98,Kuepper03}. By
Grigas \emph{et al.}~\cite{Grigas09,Grigas08}, the electronic
structure of both paraelectric and ferroelectric phases of \SPS was
calculated in the cluster approach. For the $\rm Sn_2P_2Se_6$
selenide analog, the electronic structure have been
investigated~\cite{Caracas02} by first principles calculations only
for the paraelectric phase. The electronic structure and phonon
spectra pressure dependence for acentric layered rhombohedral
crystal $\rm SnP_2S_6$ were investigated theoretically in LDA
approach~\cite{Rushchanskii06}. For this compound, the tin cations
are almost fully ionized ($\rm Sn^{4+}$ charge state) what exclude
possibility of stereochemical activity of their $5s^2$ electron lone
pair. The electronic structure of high charged $\rm (P_2S_6)^{4-}$
and $\rm (P_2Se_6)^{4-}$ anion clusters was discussed in
papers~\cite{Kuepper03,Caracas02,Rushchanskii06} at analysis of $\rm
Sn_2P_2S_6$, $\rm Sn_2P_2Se_6$ and $\rm SnP_2S_6$ electron energy
spectra. Analysis of \PS cluster chemical bounding was also done in
Hartree-Fock approach~\cite{Smirnov00}. These anions arrangement
have been investigated experimentally and theoretically in different
approximation for the layered crystals like $\rm M_2P_2S_6$ (M~--
Fe, Ni, Mn,
\ldots)~\cite{Piacentini82,Brec86,Ohno86,Sugiura96,Zhukov95}. For
the $\rm CuInP_2Se_6$ layered compound with two differently charged
cations, the SOJT effect was established as an origin of cooper
ferrielectric ordering~\cite{Fagot-Revurat03}.

In this paper, the first-principles calculations in LDA approach of
DFT for electronic structure of \SPS crystal in the paraelectric and
the ferroelectric phases were used for analysis of chemical bonds
transformation at the spontaneous polarization appearance and for
establishing of the ferroelectric state origin in the phosphorus
containing chalcogenides. As a background of the investigations, the
free \PS structure group (atomic cluster) electron spectra and
peculiarities of their molecular orbitals were considered. The
electron structure and chemical bonding nature in the paraelectric
phase will be also analyzed. The stereochemical activity of the
electron lone pair of tin cations is examined in detail. The growth
of covalence and ions recharging are related to the structure
spontaneous polarization. Finally, the influence of tin by lead and
sulfur by selenium substitution on crystals properties is discussed.

\section{Method of calculations}

The calculations of the band structure of both phases of \SPS
crystal as well as energy levels of \PS molecule has been performed
by means of the package program ABINIT~\cite{ABINIT} (total and
projected densities of states where calculated using
SIESTA~\cite{Soler02} software) in the framework of the DFT, using
the local density approximation for representing the
exchange-correlation interaction. A basis set of ~28000 plane waves,
restricted by the kinetic energy $E_{cut} = 25$~Hartree has been
used. The tin, sulphur and phosphorus atoms had the following
electron configurations: Sn: [Kr] $5s^25p^2$; S: [Ne] $3s^23p^4$;
and P: [Ne] $3s^23p^3$, respectively. The 'frozen' core electron
configurations for each atom is shown in brackets. The
first-principles pseudopotentials in the Hartwigsen-Goedecker-Hutter
scheme~\cite{Hartwigsen98} have been applied and the integration
over irreducible part of the Brillouin zone has been done by means
of the thetrahedron method using the $4\times4\times4$
Monkhorst-Pack mesh~\cite{Monkhorst76} of {\bf k}-points. The chosen
parameters were sufficient for a good convergency in the
calculations. Prior to commence the calculation of physical
properties of \SPS crystal, we carried out the structural
optimization, which minimized total energy of the system
simultaneously with the forces~\cite{Chan93} acting on atoms. The
spin-orbit interaction was not taken into account in our
calculation. The parameters of \SPS crystal obtained after
structural relaxation can be compared with experimental data,
presented in~\cite{Dittmar74,Cleary92}. The comparison of
experimental and calculated relaxed values of lattice constants
demonstrate the difference of about ~3\% in order of magnitude.

\section{Electronic structure of $\rm \mathbf{P_2S_6}$ cluster}

The molecular orbitals of \PS cluster create covalent P~---~S
and P~---~P bonds. Their hybridization with tin atomic orbitals
determines electronic structure of \SPS crystal. The electronic
energy spectra of this material could be analyzed by calculation of
free \PS cluster electronic structure with following accounting of
their molecular orbitals hybridization with atomic orbitals of tin.
\begin{figure}[b!]
\hspace{-0cm}
\includegraphics[width=1.0\columnwidth]{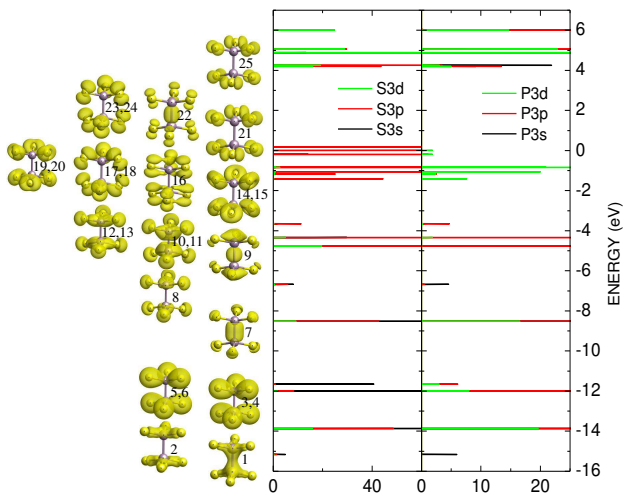}
\caption{\label{fig2} The partial densities of states and space
distribution of electron density for the molecular orbitals of \PS
cluster in free state.}
\end{figure}
The calculated energy spectrum and partial densities of electron states
for $s$, $p$ and $d$ orbitals of phosphorus and sulfur atoms of free
\PS molecule illustrate the formation of different molecular
orbitals at creation of covalent P~---~S and P~---~P bonds. The
spatial electron density distribution for related energy levels
reflects peculiarities of these bonds. It is seen that for cluster
the energy level near $-15$~eV mostly is determined by hybridization
of phosphorus $s$ orbitals. Here it have been also found some
contributions of sulfur $s$ and $p$ orbitals. The hybridization of these
atomic orbitals (scheme (1) at (Fig.~\ref{fig2})) creates bonding
P~---~P and P~---~S molecular orbitals. The level near $-14$~eV is
determined by antibonding combination of two phosphorus $s$ orbitals
and by bonding hybridization of $s$ orbitals (2) of phosphorus and
sulfur atoms.

The group of the levels near $-12$~eV mostly is formed by sulfur $s$
orbitals. These levels are related to the P~---~S bonding and
P~---~P antibonding molecular orbitals (3,4,5,6). The levels near
$-6.5$~eV and $-8.5$~eV appear as a replica of doublet of the levels
near $-15$~eV and $-14$~eV in results of antibonding hybridization
of phosphorus and sulfur atoms $s$ orbitals. The molecular orbitals
for levels near $-8.5$~eV are P~---~P bonding and P~---~S
antibonding (7), the orbitals with energy near $-6.5$~eV are
antibonding for all P~---~S and P~---~P bonds (8). In interval
between $-3.5$ and $-4.7$~eV, the energy levels are created by
bonding hybridization of phosphorus and sulfur $p$ orbitals. These
orbitals (9, 10, 11, 12, 13) are bonding for P~---~P and P~---~S
bonds. In region from 0 till $-1.5$~eV, the energy levels of cluster
are also formed by $p$ orbitals of phosphorus and sulfur atoms. They
are hybridized in P~---~P bonding and P~---~S antibonding molecular
orbitals (14~--23). Here the contribution from phosphorus $d$
orbitals is also presented.

It is worth to be noted that upmost 23rd energy level of \PS
molecule is double degenerated due to high symmetry but only half
occupied. Such peculiarity can cause instability of this complex by
means of Jahn-Teller like mechanism.

From detailed analysis of the electron energy spectrum, the important
information about creation and character of the chemical bonds in
 \PS cluster could be found (Fig.~\ref{fig2}). The bond P~---~P
is determined by $\sigma$ hybridization of phosphorus $s$ orbitals
(the level near $-15$~eV (1)) and by their replica near $-8.5$~eV
(7). In P~---~P bond, the contribution from $\pi$ hybridization of
$p_x$ and $p_y$ orbitals of phosphorus, that are oriented normally
to the bond direction, is also presented. The levels of these
orbitals (9) are placed near $-4$~eV. However, the essential
contribution into energy of the P~---~P bond adds the $\sigma$
hybridization of phosphorus $p_z$ orbitals which are oriented along
the bond. The bonding combination $\sigma\left(p_z + p_z\right)$
(22) has filled by electrons with the energy level in the range $0
\div -1.5$~eV. Also the nonbonding combination $\sigma^*\left(p_z -
p_z\right)$ which is related to the empty energy level has been
found in this energy region. Some contribution to the bond between
$\rm PS_3$ structure pyramids of \PS cluster also came from
hybridization of $p$ orbitals of sulfur atoms that belong to
different pyramids.

By $\sigma$ hybridization of phosphorus and sulfur $s$ orbitals, the
P~---~S bonds are created. Also the hybridization of $p$ orbitals of
phosphorus and sulfur atoms is observed (Fig.~\ref{fig2}). Such
hybridization has obviously both $\sigma$ and $\pi$ character. Thus
the next scheme for appearing of molecular orbitals that form
P~---~S bonds in $\rm PS_3$ structural pyramid could be proposed.
The phosphorus atom realizes $sp^2$ hybridization from which three
symmetrically oriented bonds with involving of sulfur $p$ orbitals
appear. Two electrons from surrounding cations in a crystal lattice
and one $s$ electron of phosphorus atom (excited on $d$ orbital)
supply filling of covalent P~---~S bonds.

In \SPS crystal, all energy levels for \PS clusters are
occupied, and in the ionic bonding approach, the charge states $\rm
S^{2-}$ for sulfur ions and $\rm P^{4+}$ for phosphorus ions are
expected. However, as it follows from calculations for \PS
molecular orbitals, enough high charge density is found at
phosphorus atoms. By this matter, not high positive charge
(drastically smaller than $+4$) is expected for the phosphorus ions.
In addition, the calculations provide evidence about high electronic
density at the middle of P~---~P bond.

\section{Electronic structure of $\rm \mathbf{S\lowercase{n}_2P_2S_6}$ crystal}

At building of \SPS crystal structure with two formula units in the
elementary cell (Fig.~\ref{fig1}), the quantity of energy levels of
\PS clusters is doubled with their energies splitting in results of
the inter-cluster interaction. Also, the energy levels of $5s^2$
orbitals of four $\rm Sn^{2+}$ cations are added to the structure of
the crystal valence band (VB). The cation's $5p^2$ orbitals
participate in formation of the conduction band of crystal. The
scheme of electron orbital hybridization in the crystal could be
presented as additive combination of above described scheme of \PS
cluster orbitals creation with scheme of these molecular orbitals
hybridization with tin atomic orbitals. Generally, for the crystal,
four tin atoms, four phosphorus atoms and 12 atoms of sulfur in the
elementary cell have 108 valence electrons that are placed at 54
energy levels in the VB.

\begin{figure}[t!]
\hspace{-0cm}
\includegraphics[width=1.0\columnwidth]{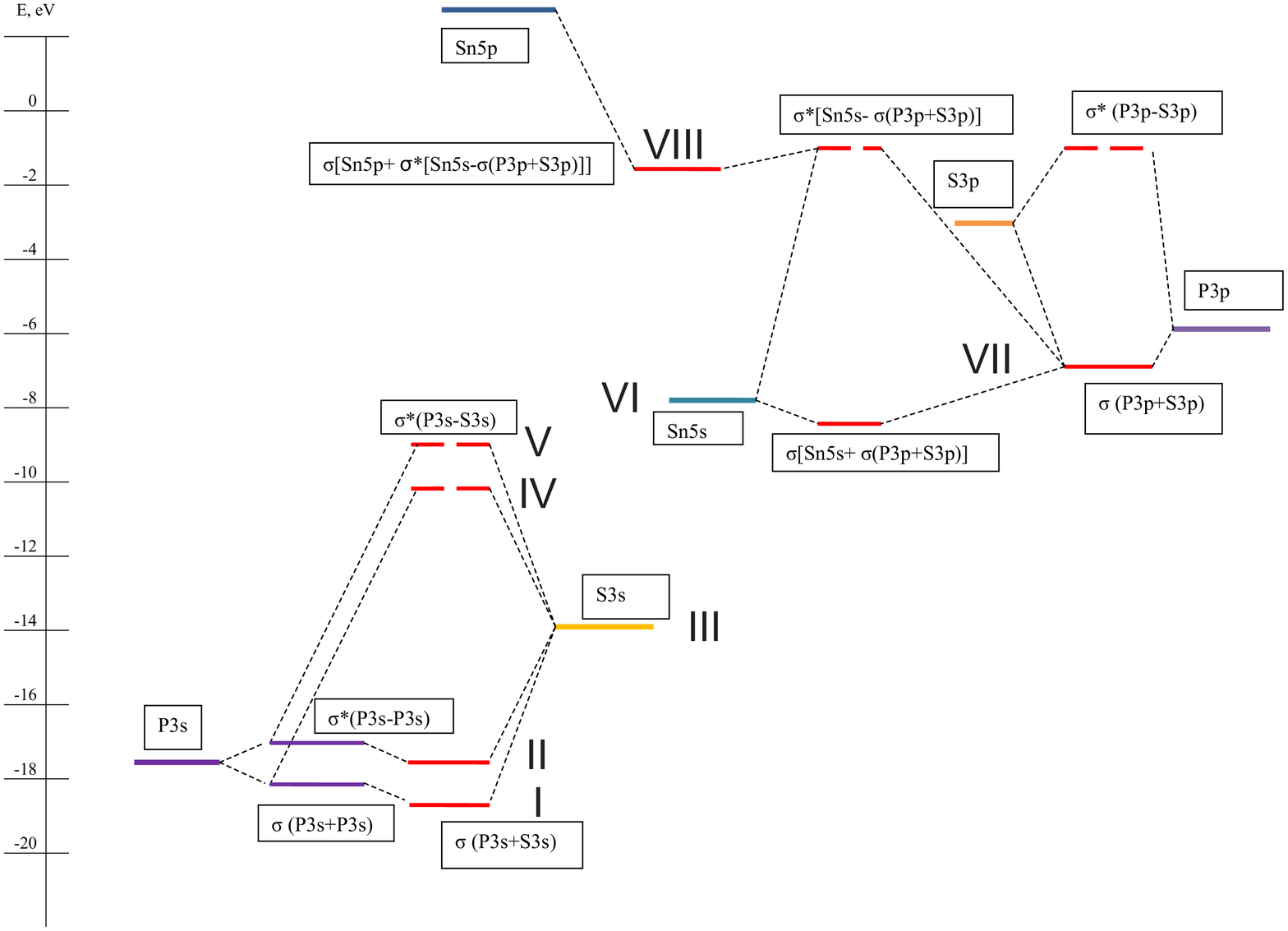}
\caption{\label{fig3} The hybridization scheme for electronic
orbitals in \SPS crystal}
\end{figure}

\begin{figure}[t!]
\hspace{-0cm}
\includegraphics[width=1.0\columnwidth]{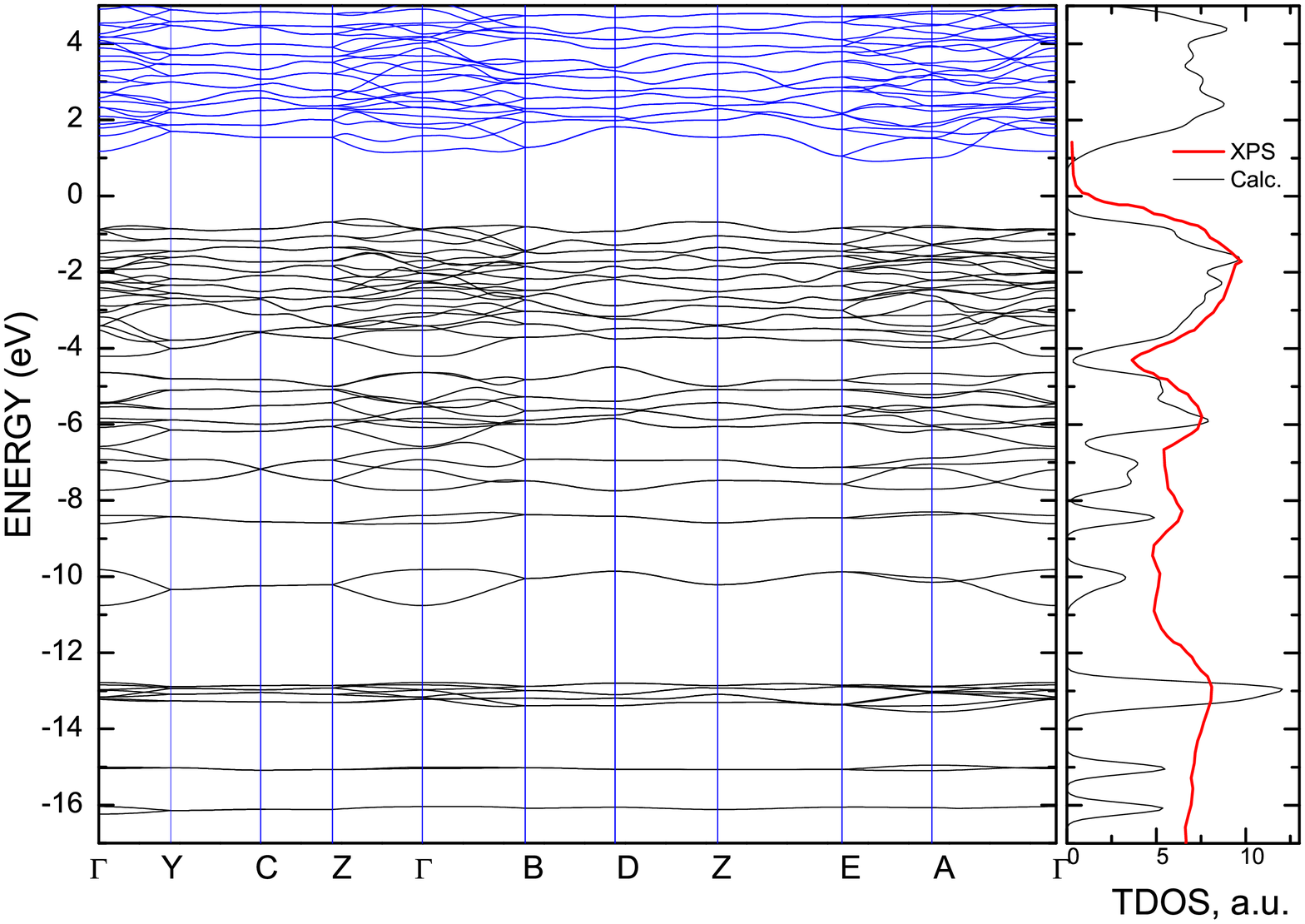}
\hfil\vspace{-0.7cm}
\includegraphics[width=1.0\columnwidth]{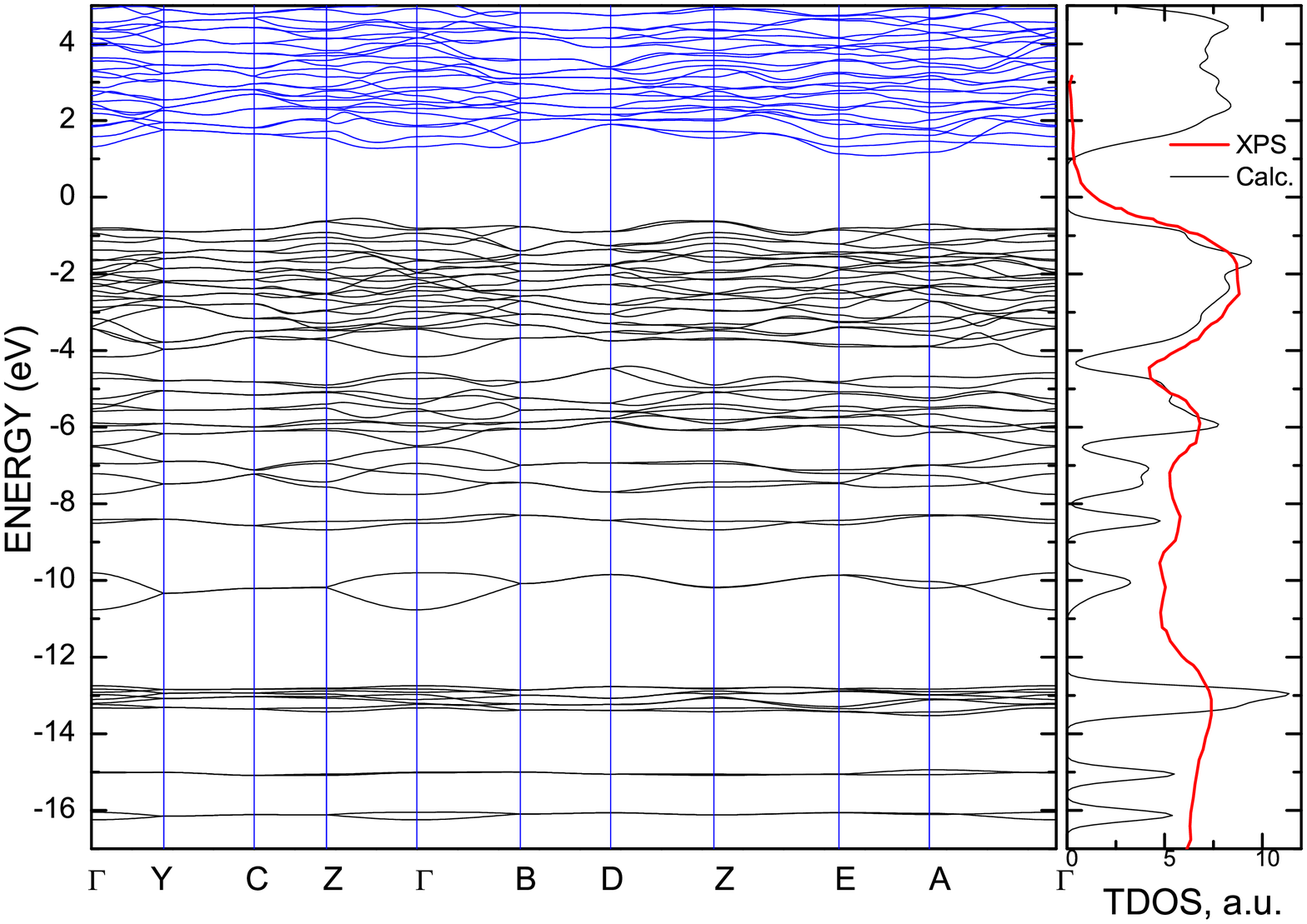}
\caption{\label{fig4}The electron energy spectrum of \SPS crystal in the
paraelectric (top) and the ferroelectric (bottom) phases. The calculated total
density of states is compared with experimental XPS
data~\cite{Grigas09}.}
\end{figure}

It's known~\cite{NIST} that energy positions of the atomic orbitals
of phosphorus, sulfur and tin are next: P\;$3p = -8.35$~eV, P\;$3s =
-17.13$~eV; S\;$3p = -10.28$~eV, S\;$3s = -20.8$~eV; Sn\;$5p =
-4$~eV, Sn\;$5s = -11$~eV. The scheme of hybridization
(Fig.~\ref{fig3}) could be proposed which is in agreement with
calculated energy spectra for free \PS cluster (Fig.~\ref{fig2}) and
for \SPS crystal (Fig.~\ref{fig4}). At the crystal structure
formation, the energy of phosphorus valence orbitals almost doesn't
change while the bonding energy for sulfur valence orbitals strongly
lowers (almost by 7~eV). This is in agreement with raised electronic
density on the sulfur anions and with enough high electronic density
surrounded the phosphorus atoms.

In accordance to the calculated energy spectra and densities of states
of \SPS crystal (Fig.~\ref{fig4}), its VB could be divided into
eight subbands which are labeled at the hybridization scheme of
atomic and molecular electronic orbitals. Remember, that for the
energy spectrum of free \PS cluster only seven subbands were found
(Fig.~\ref{fig2}). For \SPS crystal, the additional levels of tin atomic
$s$ orbitals are placed near $-8$~eV and they are related to the
sixth subband of the VB.
\begin{figure}[t!]
%\hspace{-1.7cm}
\includegraphics[width=1.0\columnwidth]{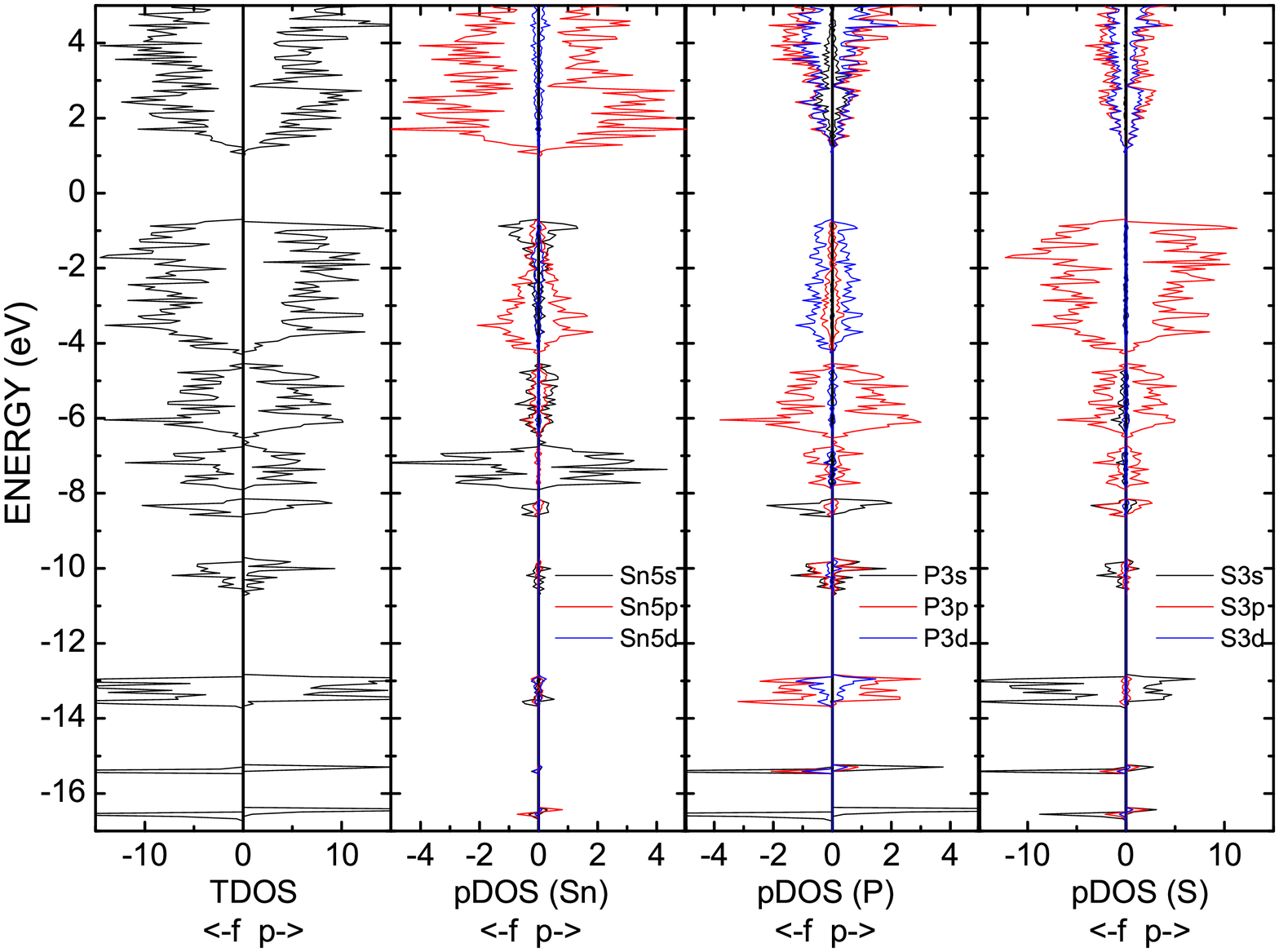}
\caption{\label{fig6}The total and partial electron densities of states
for \SPS crystal in the paraelectric (denoted by p) and
ferroelectric (denoted by f) phases.}
\end{figure}

\begin{figure*}[t!]
%\hspace{-1cm}
\includegraphics[width=2.0\columnwidth]{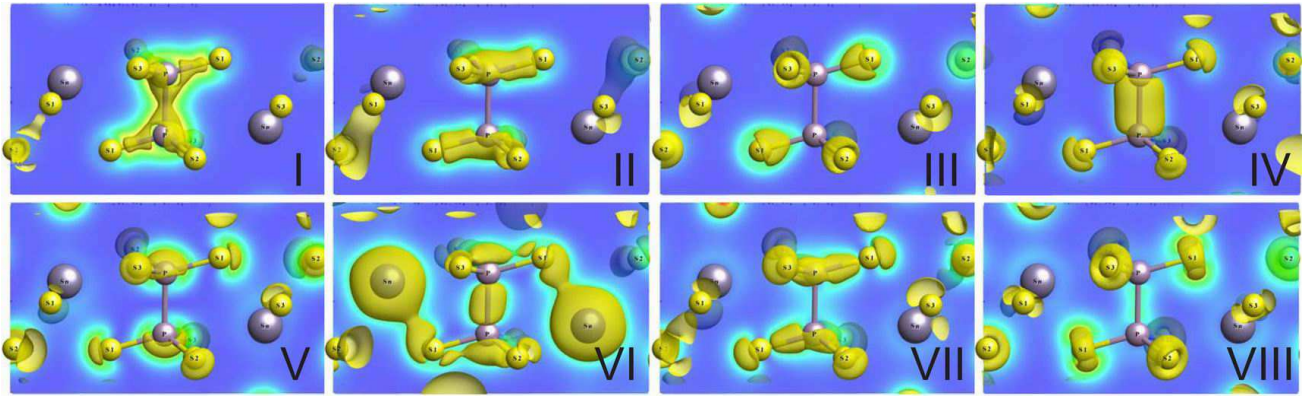}
\caption{\label{fig7}The spatial charge density distribution (in plane
which contains S~---~P~---~P~---~S bonds) for orbitals in the
valence subbands I~-- VIII for the paraelectric phase of \SPS crystal.}
\end{figure*}

The subband I contains the two energy levels near $-16.5$~eV for
which a contribution of phosphorus $s$ orbitals is dominated
(Fig.~\ref{fig6}). Here, a small contribution of sulfur $s$ orbitals
is also presented, and a minor appearance of sulfur $p$ orbitals is
observed. The contribution of tin atoms valence orbitals in this
subband is specific peculiarity of the crystal energy spectrum.
Generally, the bonding orbitals of covalent P~---~P and P~---~S
bonds are created in subband I (Fig.~\ref{fig7}).

The subband II includes the two energy levels near $-15.5$~eV. It is
formed by antibonding combination of two phosphorus atoms $s$
orbitals and by bonding hybridization of $s$ orbitals of phosphorus
and sulfur atoms. For these levels also, a some contribution from
phosphorus $d$ orbitals is observed.

The subband III in region from $-13$~eV to $-13.8$~eV has eight
energy levels for which electron charge density is mostly localized
at sulfur atoms (Fig.~\ref{fig7}). Here the contribution of
phosphorus $3p$ and $4d$ orbitals is also presented. This subband is
characterized by bonding hybridization for P~---~S and antibonding
hybridization for P~---~P covalent bonds in the crystal structure.

The subbands IV and V near $-10$~eV and $-8.5$~eV (both of them
contain two energy levels) are formed by $s$ and $p$ orbitals of
phosphorus and sulfur atoms. They are the replica of the subbands II and
I and originated from their hybridization with subband III. For the
IV-th subband, the charge is mostly localized at phosphorus atoms and
it has bonding character for the P~---~P bonds and antibonding
character for the P~---~S bonds. The subband V has antibonding
character for both P~---~P and P~---~S bonds.

The subband VI with four energy levels in range from $-8$ up to
$-6.5$~eV appears in the VB of \SPS crystal as a result of
hybridization of tin atomic electron orbitals with \PS clusters
molecular orbitals. This subband is mainly formed by tin $s$
orbitals and by $p$ orbitals of phosphorus and sulfur
(Fig.~\ref{fig6}). Charge of this subband's hybridized orbitals is
located between phosphorus atoms and around tin atoms and has
Sn~---~S and P~---~P bonding character (Fig.~\ref{fig7}).
The peculiarities of orbitals hybridization for every of four levels
from this subband will be analyzed in details later with the aim of
interatomic interactions explanation that are related to the
ferroelectric phase transition in \SPS crystal.

The subband VII located in energy region between $-6.5$ and
$-4.5$~eV and contains ten energy levels. They are formed by
phosphorus and sulfur $p$ orbitals and also include small
contribution of tin electronic orbitals. This subband has bonding
character for P~---~S and P~---~P bonds and it is antibonding for
the Sn~---~S bonds.

The VIII-th subband is situated near the top of crystal's valence
band and includes 24 energy levels in the energy range from $-4.5$ till
$-0.5$~eV. The considered subband is mostly formed by lone pairs of
sulfur $p$ orbitals, with some participation of phosphorus $p$
orbitals, and it has P~---~P bonding and P~---~S antibonding
character. Here the hybridization of $s$ and $p$ orbitals of tin
atoms what determines their stereoactivity (Fig.~\ref{fig6}) is also
reproduced. The nature of $\rm Sn^{2+}$ cations stereochemical
activity in the \SPS crystal structure will be analyzed in details
later on.

\section{Hybridization of tin atomic orbitals with
molecular orbitals of $\rm \mathbf{P_2S_6}$ clusters}

\begin{figure}[t]
\hspace{-0cm}
\includegraphics[width=.70\columnwidth]{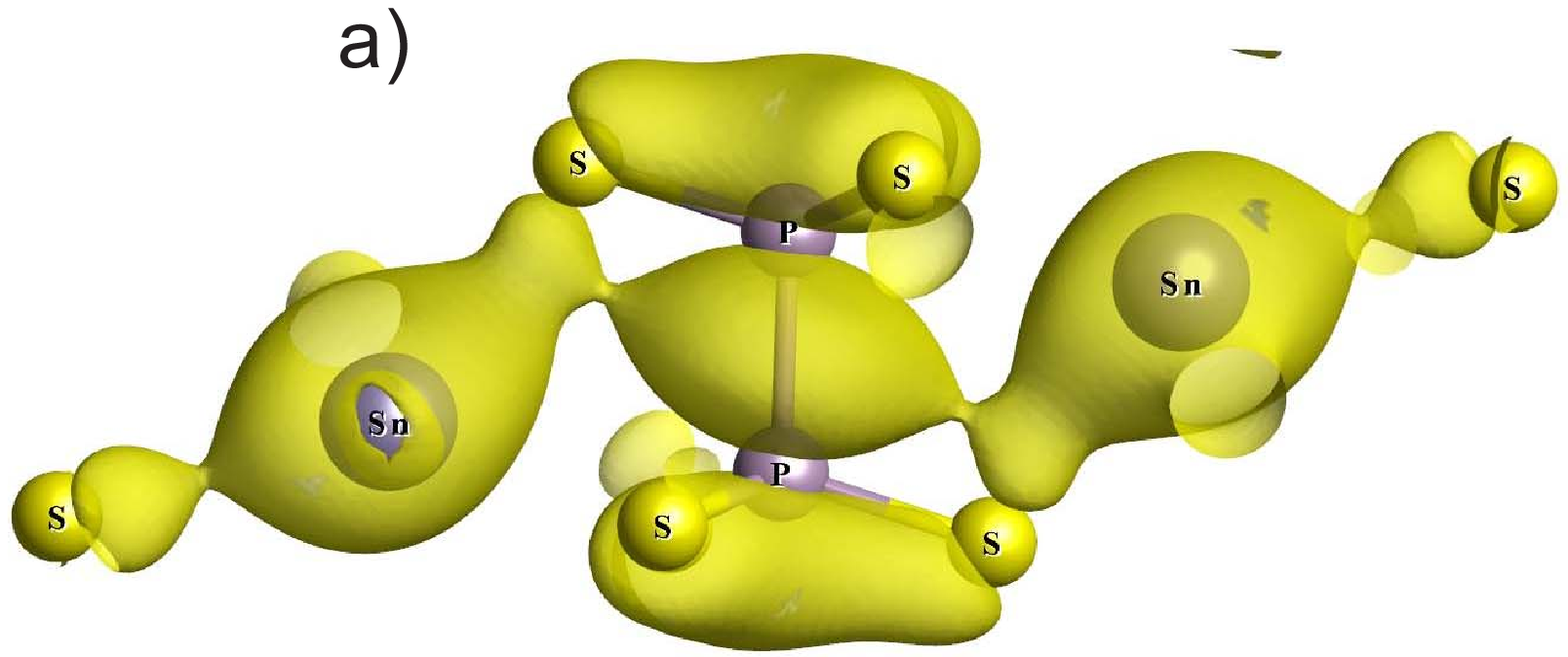}
\hfil
\includegraphics[width=.70\columnwidth]{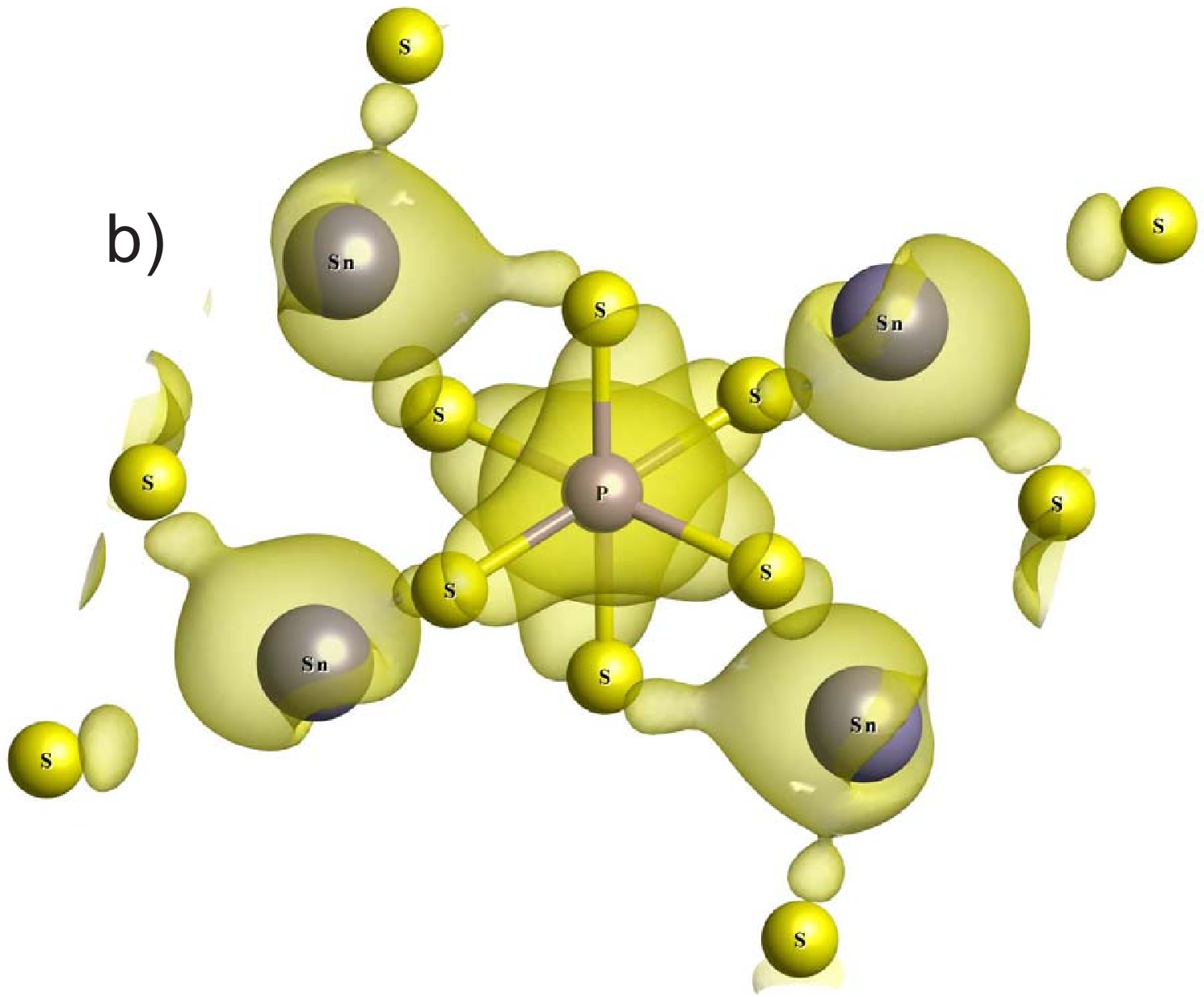}
\caption{\label{fig8}The electron density distribution~-- (a)~-- in
plane parallel to P~---~P bond, (b)~-- in perpendicular one~-- for
orbitals of 20th energy level in the valence subband VI for the
paraelectric phase of \SPS crystal.}
\end{figure}

As it was mentioned above, the valence subband VI includes four
levels in the region from $-8$ to $-6.5$~eV which are related to $s$
orbitals of tin atoms (Fig.~\ref{fig4}--\ref{fig7}). These are
levels from 17 to 20, and their spatial electronic charge
distribution illustrates a bonding peculiarities of tin atomic
orbitals with sulfur and phosphorus orbitals which create the
clusters $\rm P_2S_6$. The level 17 is characterized by enough
strong bonding of tin atoms with two nearest sulfur atoms, the
levels 18 and 19 demonstrate strong bonding of tin atom with one of
the nearest sulfur atom. The spatial charge distribution of the 20th
level is of special interest~-- here the electron density is
elongated from tin atom to the middle of P~---~P bond inside \PS
cluster (Fig.~\ref{fig8}a).

It can be easily seen that in the paraelectric phase in addition to
---~Sn~---~S~---~P~---~P~---~S~---~Sn~--- sequence of chemical bonds the
---~Sn~---~P~---~P~---~Sn~--- series also exists. Such sequence of direct bonds
of tin atoms with phosphorus atoms appears due to anisotropy of the
spatial charge distribution for the level 20 (Fig.~\ref{fig8}b). Named
distribution has a form of layers that are oriented close to the plane
$(10\overline{1})$. It is important that directions of the tin atoms
shifts at transition into ferroelectric
phase~\cite{Cleary92,Dittmar74} are also rather close to orientation
of the mentioned plane.

\begin{figure}[t]
\hspace{-0cm}
\includegraphics[width=.55\columnwidth]{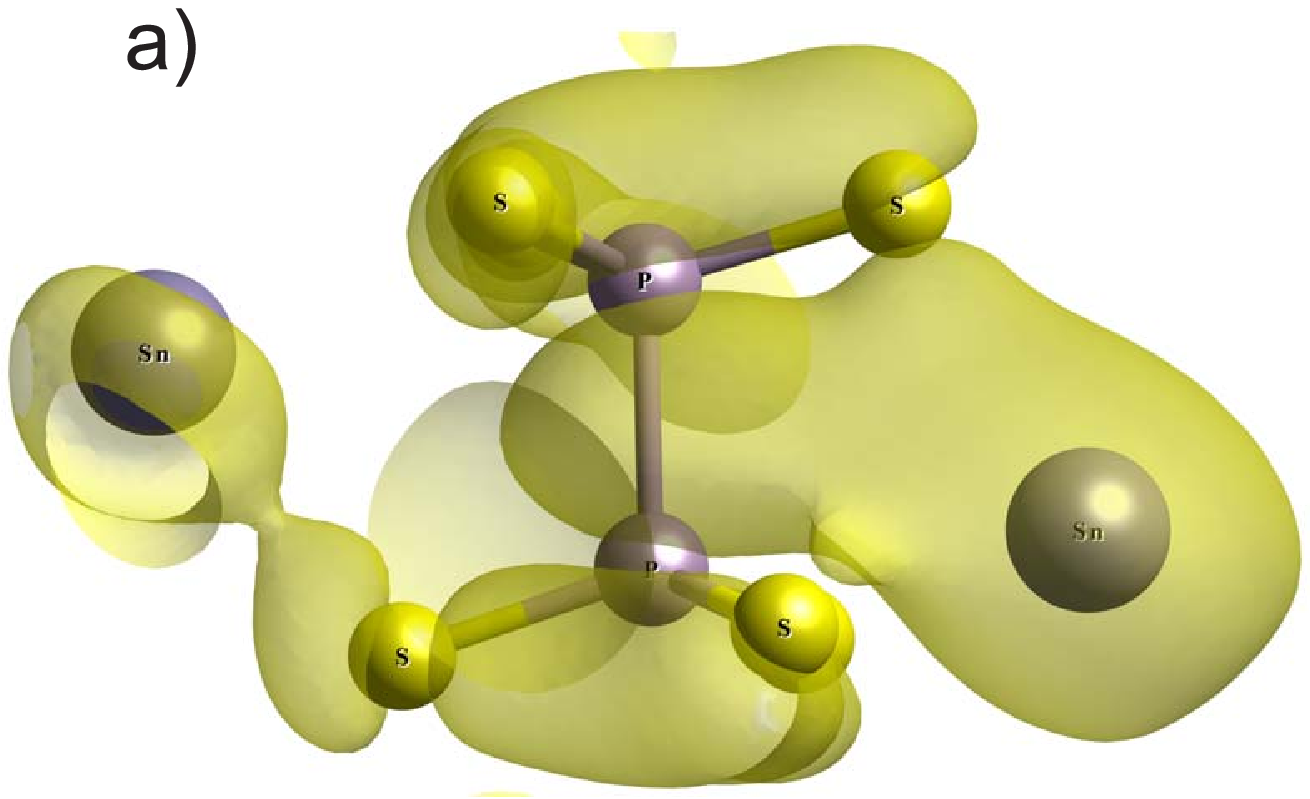}
\hfil
\includegraphics[width=.55\columnwidth]{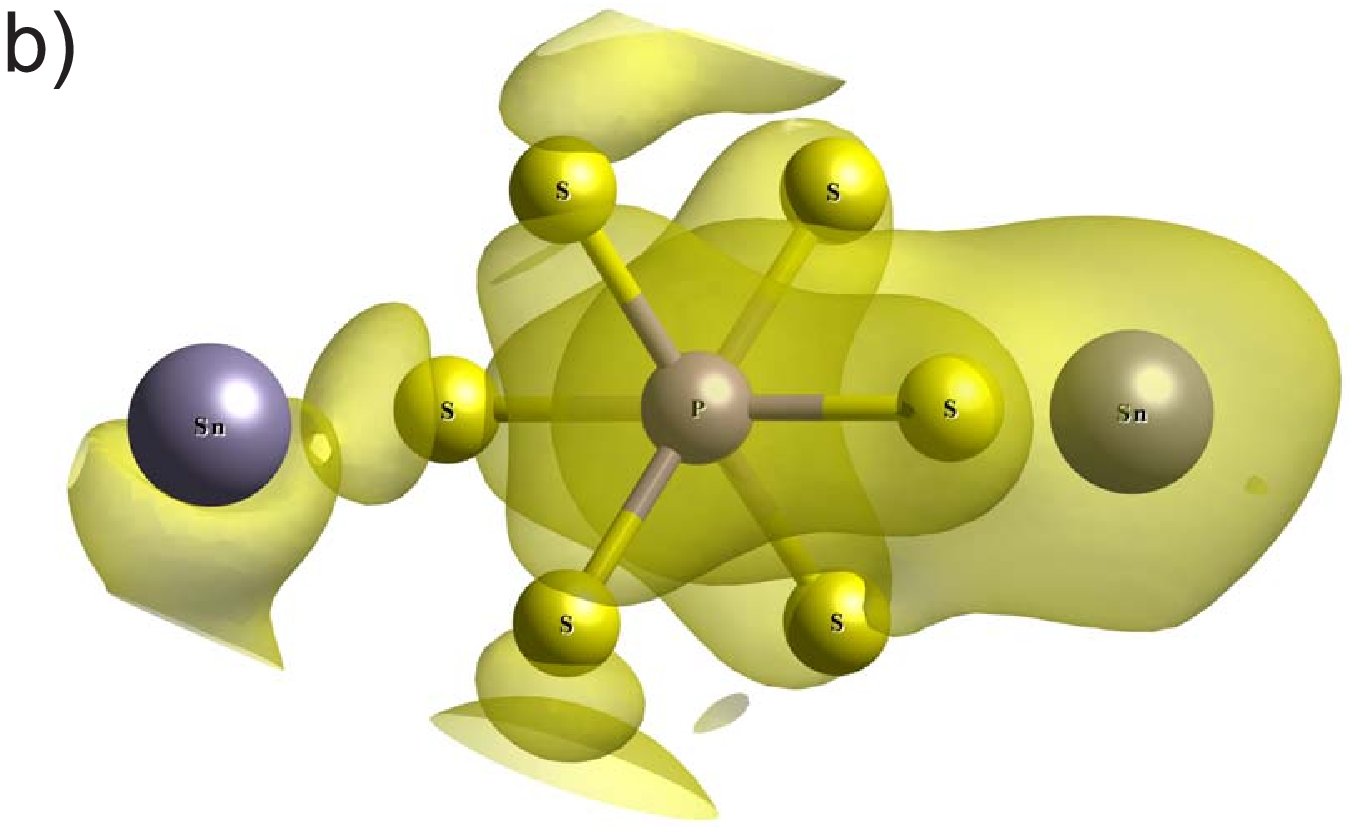}
\caption{\label{fig9}The electron charge distribution~-- (a)~-- in
plane parallel to P~---~P bond, (b)~-- in perpendicular one~-- that
illustrate the hybridization of tin and phosphorus orbitals for the
20th energy level in the valence subband VI of the \SPS
ferroelectric phase.}
\end{figure}

In the ferroelectric phase, the two pairs of nonequivalent tin atoms are
presented in the crystal structure what is clearly illustrated by
the spatial distribution of the electron charge density for the 20th
level (Fig.~\ref{fig9}a). Near mentioned $\rm P_2S_6$ cluster, one
of the tin atoms, further $\rm Sn_2$, is approached to the middle of
P~---~P bond. At this, the electron charge distribution which surrounded
the $\rm Sn_2$ atom connects with charge distribution between two
phosphorus atoms. Another tin atom ($\rm Sn_1$) goes away from the
middle of P~---~P bond, and surrounded $\rm Sn_1$ atom electron
charge distribution is oriented to one of the sulfur atoms. Such
difference in hybridization of the electron orbitals for two types
of tin atoms in the ferroelectric phase determines disappearance of
the layer-like anisotropy for the electron  spatial charge
distribution (Fig.~\ref{fig9}b).

\section{Transformation of electron energy spectra at transition from
paraelectric phase into ferroelectric phase}

According to experimental
data~\cite{Vysochanskii09,Bourdon02,Apperley93,Grigas09} the
chemical bonds and electron energy spectra of \SPS crystal have
noticeable changes at the ferroelectric phase transition. The
calculated energy spectra demonstrates the changes in the energy gap
and in positions of all energy levels of the VB~-- energies of
electron states density peaks in the VB shifts approximately by
0.5~eV (Fig.~\ref{fig4},\ref{fig6}). In acentric phase the
degeneration of electron energy levels disappears in some regions of
the Brillouin zone, what determine higher smearing of the energy
distribution of electron density of states. It is important to
mention that the lowering of the electron density of states near the
top of the VB is also observed at transition into ferroelectric
phase together with a rise of the energy gap.

\begin{figure*}[t!]
\hspace{-0cm}
\includegraphics[width=2.0\columnwidth]{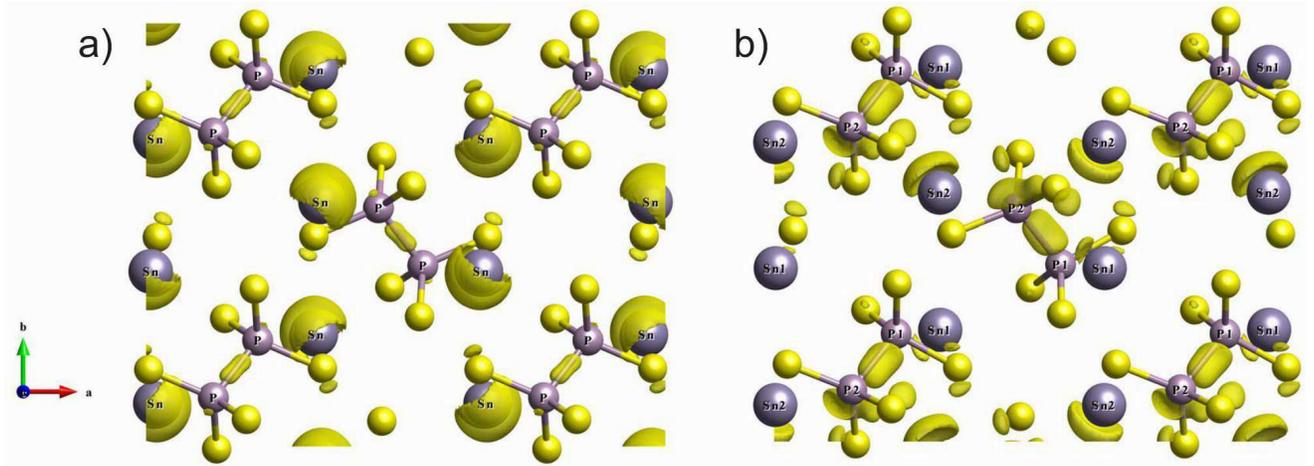}
\caption{\label{fig10}The electron density space distribution for
the VI valence subband in the paraelectric (a) and the ferroelectric (b)
phases of \SPS crystal.}
\end{figure*}

The changes of electron energy spectra obviously reflect an important
role of electron-phonon interaction in nature of the \SPS crystal
spontaneous polarization. Such interaction is illustrated by
a transformation of the spatial electron density distribution at
change of the atoms coordinates in the crystal structure. The
squares of wave functions for electron orbitals, which are summed
for the energy levels of valence subband VI in the paraelectric and
the ferroelectric phases of \SPS crystal, are shown at Fig.~\ref{fig10}.
In the elementary cell of centrosymmetric structure, a similar
distribution of electron density around four tin atoms is observed.
This distribution reflects the stereoactivity of the lone electron
pair $5s^2$ of cations $\rm Sn^{2+}$. Also, the spatial charge
distribution around phosphorus atoms is similar. This is in
agreement with the presence of inversion center at the middle of
P~---~P bond. In the acentric structure, the two pairs of tin atoms, with
different distribution of surrounding charge, appear. The nonequivalence
of the electron density distribution near the phosphorus atoms
is also seen. It should be mentioned that elevated electron density is
located in the vicinity of neighboring tin atoms $\rm Sn_2$ and
phosphorus atoms $\rm P_1$. Also, lowering of the surrounding charge
is observed for the adjacent atoms $\rm Sn_1$ and $\rm P_2$.
Acentricity of the \PS clusters is also reflected in deformation of
the charge distribution along the P~---~P bonds.

Let's analyze peculiarities of the orbitals hybridization for study
of the tin $5s^2$ electrons stereoactivity and their contribution
into lattice spontaneous polarization. In the crystal structure the
tin atoms are placed in the polyhedrons that are created by eight
sulfur atoms. At beginning we will consider the mixing of tin
orbitals with valence orbitals of surrounding sulfur atoms. Further, it
will be analyzed a role of hybridization between tin atomic orbitals
and \PS cluster molecular orbitals in change of relation between
short range and long range interactions in \SPS crystal lattice,
which induce ferroelectric phase transition.

\section{Stereoactiviy of lone $\mathbf{5s^2}$ electron pair of $\rm \mathbf{Sn^{2+}}$ cations}

By first principles calculations in approach of frozen phonons, it was
found~\cite{Rushchanskii07} that in \SPS crystal the polar optic
mode with symmetry $B_u$ in the paraelectric phase could be destabilized
only at account of their nonlinear interaction with fully
symmetrical $A_g$ optic mode. Generally, it is necessary to include the
nonlinear interaction of $A_gB_u^2$ type for all 13 $B_u$ and 15
$A_g$ normal vibrations of the lattice with symmetry $P2_1/c$ in the
paraelectric phase. Thereafter it could be then explained the variation of the
atomic coordinates at transition into the ferroelectric phase with
symmetry $Pc$~-- two tin cations (that are related by the symmetry
plane) have some shift relatively to the anion sublattice, and another
two tin cations found a bigger shift (flipping) relatively to their
positions in the paraelectric phase. It was determined, that effective
potential in the normal $A_g$--$B_u$ coordinates has three minima~--
central one reflects the metastable paraelectric state, and two side
minima are related to two domains of the ferroelectric phase in uniaxial
ferroelectrics.

\begin{figure}[t]
\hspace{-0cm}
\includegraphics[width=0.90\columnwidth]{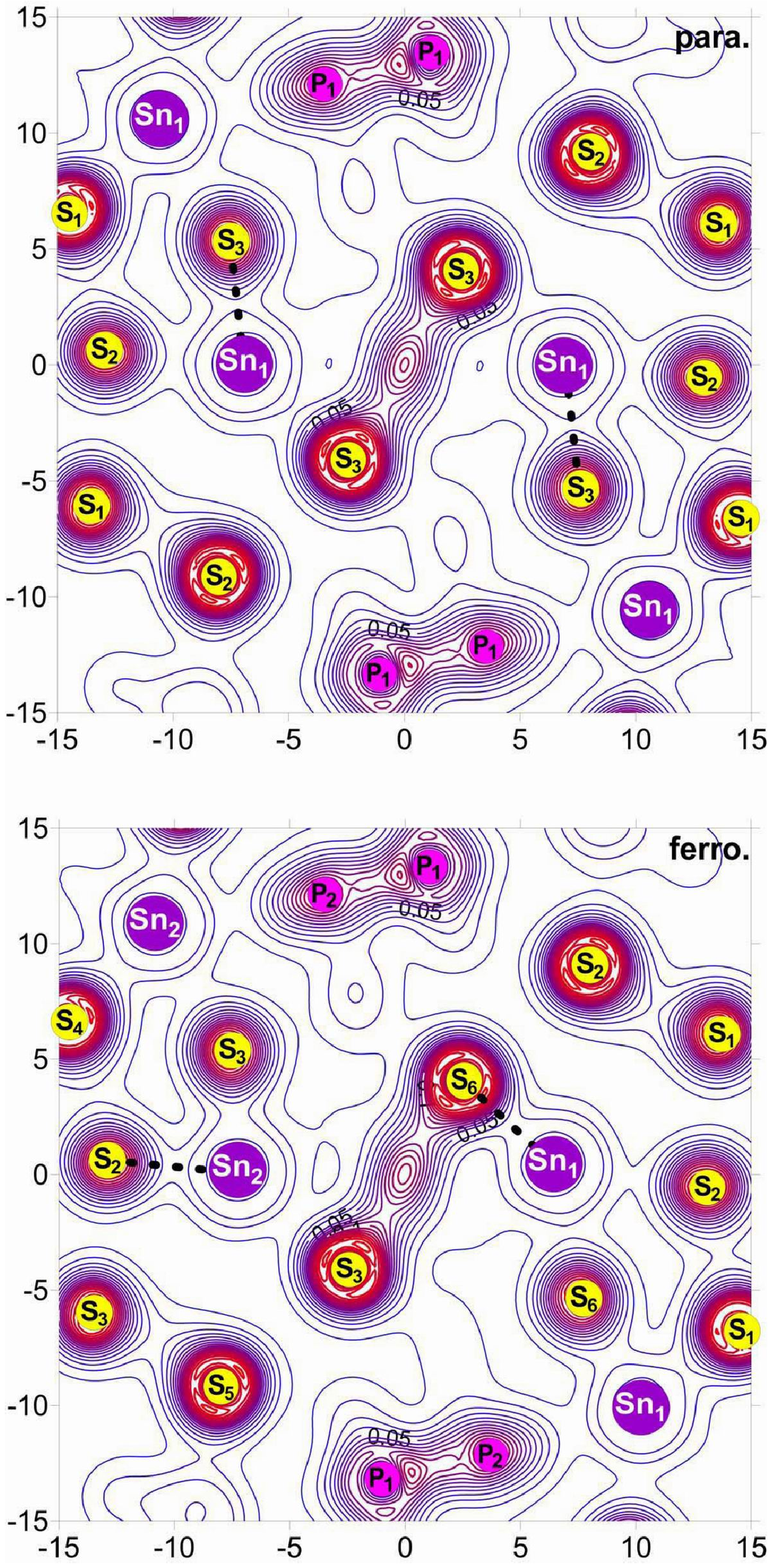}
\caption{\label{fig11}The transformation of the electron density space
distribution in the vicinity of tin cations at the ferroelectric phase
transition in $\rm Sn_2P_2S_6$ crystal. The strongest Sn~---~S bonds is
shown by dashed lines.}
\end{figure}

The complicate potential relief for \SPS crystal is obviously
determined by relaxation of electron lone pair of tin cations. The
electron pair in $5s^2$ configuration is stereoactive~-- $s$
orbitals of cations are hybridized with $p$ orbitals of sulfur. The
antibonding energy level as a result of such hybridization is still
occupied by electrons. For the energy gain, this level interacts
with tin $p$ orbitals~-- $sp^2$ hybridization is realized for which
the bonding level lowers their energy and antibonding level elevates
higher of Fermi level. Such hybridization is proportional to the
acentricity of surrounding crystal field, and it induces deformation
of built by eight sulfur atoms dodecahedron which surround the tin
cation. Thereafter the metal cation shifts away from the
dodecahedron center, and space distribution of the electron charge
is clearly different from spherical. The stereoactivity of electron
lone pair represents the second order Jahn-Teller effect~-- SOJT.

Already in the paraelectric phase, the tin cations have placed in general
position of the elementary cell and their surrounding by electron
density is definitely not spherical~-- almost rigid dipoles exist
which are not aligned. At cooling the stereoactivity of electron
lone pair growths and the orientation of cation shifts in nearest
elementary cells is correlated by dipole-dipole interaction. The
spontaneous polarization appears with two contributions~-- ''displacive''
and ''order/disorder''.

The experimental data of XPS spectroscopy~\cite{Grigas09,Grigas08}
about electron energy spectra near the top of VB in the paraelectric and
the ferroelectric phases of \SPS crystal confirm the lowering of the
electron energy states density at transition into acentric
structure.  The calculations of energy spectra show big enough
contribution of tin $s$ orbitals into electron states density near
the top of the VB and permit a possibility for tracing of their change at
the phase transition (Fig.~\ref{fig6}). The calculated space distribution of
total electron density illustrates an evolution of electron lone pair
at transition from the paraelectric phase into the ferroelectric one
(Fig.~\ref{fig11}). The appearance of the tin cations
nonequivalence is observed: two of them shifts in a direction of those
sulfur atoms with which in the paraelectric phase they have the biggest
overlap of electron orbitals; alternative two cations ''flip''
the biggest overlap of their electron orbitals in direction of other
sulfur atoms (Fig~\ref{fig11}).

The high coordination of tin cations could be obviously described by
taking into account their $d$ orbitals. For symmetry requirements
the hybridization of $sp^2d^5$ type satisfies and they could be
related to the positioning of the tin cations inside of the eight
caped polyhedron of sulfur anions.  Indeed, the orbitals $s$, $p_x$,
$p_z$, $d_{xz}$, $d_{xy}$, $d_{zy}$, $d_{z^2}$, $d_{z^2-x^2}$
transform on the irreducible presentations $A_g$, $B_u$, $B_u$,
$A_g$, $B_g$, $B_g$, $A_g$, $A_g$ (orientation of the monoclinic
symmetry plane coincides with cartesian plane XZ) that have fully
symmetrical combination.

\begin{figure}[t]
\hspace{-0cm}
\includegraphics[width=.95\columnwidth]{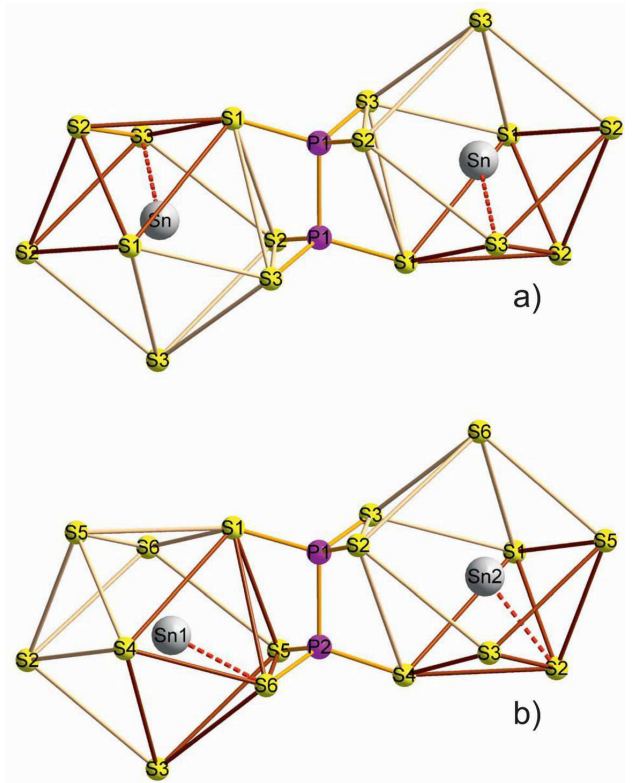}
\caption{\label{fig12}The transformation of sulfur atoms coordination
polyhedrons around the tin cations at the ferroelectric phase transition
in $\rm Sn_2P_2S_6$ crystal. The sulfur atoms that have stronger bonds with
tin atoms are linked by dark lines. The strongest Sn~---~S bonds is
shown by dashed lines.}
\end{figure}

In the elementary cell of the paraelectric phase, the four such dodecahedrons
are equivalent~-- they pair-by-pair are related by second order
screw axis or by glade mirror plane (Fig.~\ref{fig12}a). In the
ferroelectric phase as a result of earlier discussed charge density
redistribution, the inversion center and symmetry axis disappear.
The two pairs of nonequivalent dodecahedrons appear. In one type of the
dodecahedrons at the phase transition, the electron density switches
between two nearest sulphur atoms which corresponds to the strongest
Sn~---~S bond. In the other type of dodecahedrons at cooling from the
centersymmetric phase to acentric one, the ''flipping'' of  the
electron density between almost oppositely oriented Sn~---~S bonds
is observed (Fig.~\ref{fig12}b).

The built by sulfur atoms pyramids could be divided in the
coordination polyhedrons (Fig.~\ref{fig12}). Localized at hybridized
$sp^2$-like orbitals of tin atoms, the electron density is oriented in
direction opposite to the base of named pyramids. This is a direction
to the three sulfur atoms with weaker Sn~---~S bonds.

A characterization of the chemical bonds changes at phase transition
could be found by compare of calculated data (Mulliken charges and
overleap parameters for the electron orbitals of neighbor atoms)
with experimental structure data, with M\"{o}ssbauer, XPS and NMR
spectroscopies data. For the paraelectric phase of \SPS crystal, the
following electron configurations were calculated: for four
equivalence tin atoms Sn~-- $5s^{1.865}5p^{1.153}5d^{0.222}$; for
four equivalence phosphorus atoms P~--
$3s^{1.187}3p^{2.508}3d^{1.172}$; for one of the three types of
sulfur atoms S~-- $3s^{1.833}3p^{4.231}3d^{0.179}$. It could be seen
that $d$ orbitals are populated, what agrees with the explanation of
high dodecahedral coordination of tin cations on background of
$sp^2d^5$ type hybridization. The highest population of $d$ orbitals
is appeared for phosphorus atoms. As was mentioned earlier, for the
\PS clusters the bonds in $\rm PS_3$ structure pyramids are
determined by $sp^2$ hybridization of phosphorus atomic orbitals
with their further $\sigma$ hybridization with sulfur $p$ orbitals.
Thereafter one of two $3s^2$ electrons of phosphorus is involved
into $sp^2$ hybridization of atomic orbitals, other electron is
excited on the atomic $d$ orbital. The named $sp^2$ hybridization
involves two $p$ electrons of phosphorus, third $p$ electron by
$\sigma(p - p)$ hybridization creates covalence P~---~P bond.

In the ferroelectric phase, the electron configurations for two types of
tin atoms are the following: $\rm Sn_1$~-- $5s^{1.851}5p^{1.163}5d^{0.224}$;
$\rm Sn_2$~-- $5s^{1.850}5p^{1.177}5d^{0.226}$. It is seen that at
transition into the ferroelectric phase, the quantity of $s$ electrons lowers
by $\Delta n_s = 0.014$. Such calculations data are in agreement
with observed lowering of isomer shift for spectral line of
$\rm{}^{119}Sn$ M\"{o}ssbauer effect at cooling from the paraelectric
phase into the ferroelectric one~\cite{Vysochanskii09}.

The diminishing of $s$ electron quantity in the ferroelectric phase for two
types of tin atoms is compensated in a different step (level) by
growing of $p$ orbitals occupancy. Hence the total charge of tin
atoms changes from 3.240e in the paraelectric phase to 3.239e ($\rm
Sn_1$) and 3.251e ($\rm Sn_2$) in the ferroelectric phase. The calculated
rise of electron density in the vicinity of $\rm Sn_2$ atoms is in
agreement with decreasing of resonance frequency in NMR spectrum for
$\rm{}^{119}Sn$ from $-781.3$~ppm in the paraelectric phase to
$-782.5$~ppm in the ferroelectric phase~\cite{Apperley93}. The increase of the
resonance frequency till $-754$.5~ppm for $\rm Sn_1$ is obviously
determined by a occupancy redistribution from $s$ orbitals to $p$
orbitals at some lowering of total charge.

Determined by positions of XPS spectral lines~\cite{Grigas09}, the energy
of chemical binding for the $4d$ core orbitals in the ferroelectric
phase differently increases for the two types of tin cations. This is
obviously defined by growing of $5s^2$ electrons lone pair
stereoactivity and by these electrons redistribution on more
distanced from tin cores $p$ orbitals.

The changes in \SPS crystal structure at the spontaneous polarization
appearance are characterized by the calculated values for the overleap
parameters of electron orbitals of tin atoms and the nearest sulfur
atoms. In the paraelectric phase such parameter with the biggest value about 0.054
has been found for the shortest bonds Sn~---~S in the dodecahedrons
of sulfur atoms (Fig.~\ref{fig12}). It must be mentioned that in
the nearest dodecahedrons, which are related by a second order screw
symmetry axis, such Sn~---~S bonds have opposite orientations of
their projections onto plane $(010)$, which contains the spontaneous
polarization. In the ferroelectric phase the nonequivalence of two pairs of
cations $\rm Sn_1$ and $\rm Sn_2$ and surrounded them dodecahedrons
of sulfur ions appear. Here the biggest overleap parameter for
atomic orbitals (0.096) was found for the $\rm Sn_2$~---~S bond.
This bond was strong already in the paraelectric phase and it is in
addition enforced at spontaneous polarization appearance~-- the atom
$\rm Sn_2$ in the ferroelectric phase is shifted in the direction of sulfur atoms
with the biggest content of chemical bonds covalency. For other
coordination dodecahedron the strongest chemical bond $\rm
Sn_1$~---~S has the overlap parameter 0.087. Here at transition from
the paraelectric phase into the ferroelectric one, the ''flipping'' of maxima
in the space distribution of electron density occurs between bonds
of central metal atom and ligand atoms in the coordination
polyhedron. As a result of such ''flipping'' in both types of the
dodecahedron, the strongest bonds $\rm Sn_1$~---~S and $\rm
Sn_2$~---~S have nearly oriented projections onto plane $(010)$
(Fig.~\ref{fig12}b).

The changes in the electron density space distribution correlate
with variations of the interatomic distances. For example, in
the paraelectric phase the Sn~---~S bonds with the biggest overleap
parameter (0.054) are strongest and they have the smallest length in the
ferroelectric phase. For the dodecahedrons with tin atoms of $\rm
Sn_2$ type, the overleap parameter for strongest bonds increases to
0.096 and their length decreases by 0.21~\AA. Hence the occupancy
of XZ plane oriented $p$ orbitals raises~-- from the calculations
follows: $\Delta p_x = 0.048$, $\Delta p_y = -0.002$,
$\Delta p_z = 0.022$. For the dodecahedrons with tin atoms of $\rm Sn_1$ type
at the electron charge density ''flipping'' on almost
oppositely oriented Sn~---~S bond, the overleap parameter, for the
strongest bond in the paraelectric phase, lowers from 0.054 till 0.022
and  length of this bond growths by 0.26~\AA. The strongest
$\rm Sn_1$~---~S bond in the ferroelectric phase is characterized by
overleap parameter 0.087 and its length decreases by 0.35~\AA. Here
the changes for occupancies of tin $p$ orbitals are: $\Delta
p_x = 0.025$, $\Delta p_y = 0.036$, $\Delta p_z = 0.010$. We could
see that cations of $\rm Sn_2$ type donate the biggest electronic
contribution into the spontaneous polarization and this contribution has
''displacive'' character. For the cations of $\rm Sn_1$ type, the
electronic contribution is a little smaller and this one has
''ordering'' character.

The calculated changes of Mulliken charges and atomic orbitals
overleap parameters coincide with the temperature dependence of the
resonance frequencies in NMR spectra for $\rm {}^{31}P$
phosphorus~\cite{Bourdon02,Apperley93}. In the paraelectric phase
all phosphorus atoms are equivalent (calculated charge is 4.869e)
and here only one NMR line with frequency 92.12~ppm is observed. At
the spontaneous polarization appearance, the inversion center
vanishes what is associated with growth of phosphorus atoms
nonequivalence in $\rm P_2S_6$ clusters. For two types of phosphorus
atoms the following electron configurations were calculated: $\rm
P_1$~-- $3s^{1.1909}3p^{2.499}3d^{1.183}$; $\rm P_2$~--
$3s^{1.182}3p^{2.509}3d^{1.193}$. For atoms of $\rm P_1$ type with
grown calculated charge (3.251e) in result of shielding effect, the
NMR specter resonance frequency decreases to 89.2~ppm. This is in
agreement with closeness of $\rm P_1$ type atoms and $\rm Sn_2$
cations for which also a lowering of the NMR resonance frequency is
observed~\cite{Apperley93} in the result of electron density growing
in their vicinity at cooling into the ferroelectric phase. For atoms
of $\rm P_2$ type in the ferroelectric phase, the resonance
frequency rises till 93.7~ppm. In this case the weakening of the
charge shielding effect is obviously determined by diminishing of
the $s$ orbitals occupancy. By the way, for neighbor $\rm Sn_1$
cation the quantity of $s$ electrons also decreases, what induces
growth of related resonance frequency in $\rm{}^{119}Sn$ NMR
spectrum~\cite{Apperley93}.

\section{Discussion of results}

The calculated electron energy spectra, densities of electron states and
their variation at transition from the paraelectric phase into
the ferroelectric one coincide with the available structure data and
results of experimental investigation of the chemical bonds nature.
On this ground the next generalized description of sources of the
spontaneous polarization appearance in \SPS crystal could be
proposed.

The ferroelectric distortion is proportional to the difference
between the $\rm Sn_1$ and $\rm Sn_2$ positions. For this distortion
the short range (mostly Sn~---~S) repulsions must be sufficiently
small in order to allow to shift the equilibrium Sn positions from
the center of the chalcogen dodecahedron. The effective charge of
phosphorus has to be sufficiently small in order to allow the shift
of the Sn cations in the direction of  P~---~P bond. Such
requirement could be satisfied by the following way. At fully
symmetrical $A_g$ lattice vibration, the important changes of electron
charge distribution in the elementary cell occur~-- the charge
partially is waded from anions \PS onto cations Sn. Hence the $p$
orbitals of tin cations have to be occupied~-- the stereoactivity of
valence electrons of these cations is realized by their partial
hybridization with $p$ orbitals of neighbor sulfur atoms. Such
hybridization lowers the short range repulsion between tin cations
and the nearest sulfur atoms what govern their approaching. The charge
transferring between tin and sulfur atoms manages some lowering of
electrostatic interactions energy.

Generally, in the ground state (at 0~K) the metastable center-symmetric
\SPS structure is possible, for which both opposite tin atoms are a
little approached to the middle of the P~---~P bond. However, at low
temperatures the acentric structure is energetically more favorable.
At approaching of two tin cations, that are related by inversion
center at middle of P~---~P bond, their Coulomb repulsion increases
(such repulsion between cations of tin and phosphorus obviously
don't play important role because positive effective charge of
phosphorus is not big). Hence energetically more advantageous
could be approach of one tin cation to the center of \PS cluster at
the repulsion of opposite tin cation. Indeed, in the center-symmetric
structure both tin atoms are placed on distance 3.633~\AA~ far from
middle of P~---~P bond, in acentric structure such distances equal to
3.463~\AA~ for $\rm Sn_2$ atom and 3.857~\AA~ for $\rm Sn_1$ atom.
The space between considered tin atoms increases from 7.266~\AA~ in the
paraelectric phase till 7.310~\AA~ in the ferroelectric
phase~\cite{Cleary92,Dittmar74}.

Removing of $\rm Sn_1$ atom away from \PS cluster decreases
hybridization of his valence electron orbitals with molecular
orbitals of cluster. By this matter the electron charge in cluster
moves onto $\rm PS_3$ structural pyramid with $\rm P_1$ atom at
their top that is the nearest to $\rm Sn_2$ atom. At repulsion of $\rm
Sn_1$ atom from the cluster, an important change of the hybridization
character occurs, which is accompanied by localization of valence
electrons near the tin ion core and by growth of their kinetic
energy. Such processes have obviously activation character and they
determine presence of the energy barrier between central and side
minima in the three-well potential.

By such way, the induced by the fully symmetrical $A_g$ vibration,
the important changes of charge gradient in the elementary cell
determine variation of the electron configuration for the ions of
crystal lattice. The reconstruction of electron configuration modifies
the balance of interatomic interactions what induce instability of $B_u$
polar lattice vibration. Exactly by such manner, the mechanism
of lattice modes $A_gB_u^2$ nonlinear interaction could be
presented, and this one governs the tree-well potential presence for
fluctuations of the order parameter of the ferroelectric phase
transition in \SPS crystal.

The value of energy barrier in the three-well
potential~\cite{Rushchanskii07} equals near 0.015~eV, and the energy
difference between central and side minima is near 0.01~eV. Such
energetic characteristics are in agreement with our calculation of
the electron energy spectra of \SPS crystal in the paraelectric and
the ferroelectric phases. Thus, for the ferroelectric phase the full
energy was found by 0.0078~eV smaller in compare with calculated
full energy for the paraelectric phase. It must be mentioned that at
the crystal symmetry lowering the negative contribution of coulomb
interactions into full energy decreases.

The appearance of spontaneous polarization in ferroelectric crystal is
determined by variation of the chemical bonds covalency and by
delicate balance between short range repulsion forces, which
determine the relief of local potential for the phase transition
order parameter, and long range displace forces that define energy
of intercell interaction.

The hybridization of tin and sulfur atomic orbitals
which defines the appearance of ''partially rigid'' electric dipoles
(pseudospins) in result of tin cations valence electrons
stereoactivity was early analyzed. Such hybridization could be described as $sp^2$ or
$sp^2d^5$ combination of tin and sulfur atomic orbitals. This fact is
clearly demonstrated by presence of enough high density of Sn $5s$
states near the top of VB. In addition to that, it have been found
neediness of accounting for hybridization between tin atomic
orbitals and \PS molecular orbitals what has argument by presence of
$5s$ and $5p$ states of tin even at the bottom of VB~-- in the
energy range near $-17.5$~eV where the $s$ states of phosphorus
atoms are dominated.

The obtained pictures of the electron density space distribution
show (Fig.~\ref{fig8}) the presence of
---~Sn~---~S~---~P~---~P~---~S~---~Sn~--- short-range bonds chains in the paraelectric phase.
The occurrence of the short-range interactions in chains of
---~Sn~---~P~---~P~---~Sn~--- type also have been demonstrated.  Such
sequences of the short-range interactions together with long-distant
coulomb interactions determine the mean-field, which induce
correlation of the pseudo-spins at lowering of disordering influence
of heat energy. At temperature down till 0~K, the chains of central symmetric structure groups
---~Sn~---~$\rm P_2S_6$~---~Sn~---, which are related to the pseudo-spins
position in the central well of local potential, could obviously
exist also. However, the correlated ordering of the structure
motives (''dimers'') like ---~Sn~---~$\rm P_2S_6$~---, which
responses to the pseudo-spins standing in one of the side well of
local potential, is energetically more favorable.

It is important to remark about rise of anisotropy of tin and
phosphorus atoms surround in the ferroelectric phase
(Fig.~\ref{fig9}) what support increasing of their dynamic or Born
effective charges. The dynamical transfers of charge are expected to
be larger when such a hybridization involves $d$ states, for which the
interactions parameters with other orbitals are particularly
sensitive to the interatomic distance~\cite{Ederer10}. Also, the
amplitude of Born effective charges is not monitored by a particular
interatomic distance but is dependent on the anisotropy of the Sn
environment along the
---~Sn~---~S~---~P~---~P~---~S~---~Sn~---  chains. In the
paraelectric phase, the S $3p$ electrons are obviously widely
delocalized and dynamical transfer of charge can propagate along the
---~Sn~---~S~---~P~---~P~---~S~---~Sn~--- chains. In the ferroelectric phase,
these chains behave as a sequence of ''dimers''
---~Sn~---~S~---~P~---~P~---~S~---~\ldots~
for which the electrons are less polarizable.

The anomalously large dynamical charges produce big LO-TO splitting
for the ferroelectric soft phonon mode~\cite{Waghmare03}. Moreover,
this feature is associated with the existence of an anomalously
large destabilizing dipole-dipole interaction, sufficient to
compensate the stabilizing short range forces and induce the
ferroelectric instability. In materials where polar soft modes play
a major role, the dynamical charge relate the electronic and
structural properties~\cite{Zhong94}. However, for the \SPS crystal
the big LO-TO splitting for polar modes wasn't observed in phonon
spectra~\cite{Gomonnai82,Hlinka02}. Here, at 4.2~K such splitting is
in the range of $2 \div 7$~$\rm cm^{-1}$. At heating to the
temperature of phase transition in the ferroelectric phase, the
LO-TO splitting for the lowest energy optic mode of $B_u$ symmetry
(soft mode) reaches only the value of 10~$\rm cm^{-1}$.

The low frequency dielectric susceptibility temperature anomaly in
\SPS crystal don't described only by dielectric contribution of the polar
lattice vibrations. On the data of dielectric
spectroscopy~\cite{Volkov83} in the paraelectric phase, the dielectric
contribution from polar lattice vabrations into static dielectric
susceptibility reaches only near ten percent. Obviously in the range of
phonon frequencies, a significant destabilizing
dipole-dipole interaction doesn't appear. The essential contribution into dielectric
anomaly appears at frequency lowering into submillimeter diapason~--
here the relaxational dispersion have been observed~\cite{Grigas88},
which is obviously determined by nonlinear dynamic excitations.

The above attention was accented on the mixing between states of tin
valence electrons and orbitals of phosphorus and sulfur atoms across
all energy range of \SPS crystal VB. In addition, the enough large
density of phosphorus $s$ and $p$ states is presented near the top
of VB also. Such phosphorus orbitals also create the conductivity
band. The defined facts give evidence about enough strong mixing of
diffusive $s$ orbitals in structure of \SPS crystal. Obviously, the
effective occupation of phosphorus $d$ orbitals, which commonly are
enough localized, also give evidence of their important role in the
mechanism of electron-phonon interaction for this ferroelectrics.

For the $\rm P_2S_6$ structure cluster, with symmetry $D_{3d}$ in
free state, the $3s$ electron orbitals of phosphorus atoms, which
are placed at bottom of the VB (near $-15$~eV) and their bonding
$\sigma(p_z - p_z)$ orbitals, which create P~---~P bonds and have
energy level near the top of VB (in range $-1.5 \div 0$~eV), satisfy
the transformation according the $A_g$ irreducible presentation. In
addition to these orbitals of free anion cluster, the orbitals with
energies near $-8.5$ and $-4$~eV are also involved at formation of
P~---~P bond (Fig.~\ref{fig2}).

In the \SPS structure the symmetry of \PS anion clusters is lowered,
however possibility of effective mixing for the wave functions of
$s$ and $p$ orbitals of phosphorus atoms, which have identical
symmetry, is obvious. To mentioned series of four sets of P~---~P
bonding orbitals, which create high electron density at middle of
this bond (Fig.~\ref{fig7}), the combination of tin atomic orbital
related to the energy level 20 in VI subband (in range from $-8$
till $-6.5$~eV) is added. Obviously, in result of such hybridization
the similar changes at the top and bottom of the VB are observed at
appearance of spontaneous polarization in the crystal
(Fig.~\ref{fig6}).

Generally, the high effective charge and large polarizability of \PS anionic
clusters together with stereoactivity of tin electron lone pair
determine large electronic contribution from all atoms into
spontaneous polarization of \SPS crystal. Such situation is in
agreement with earlier founded~\cite{Rushchanskii07} involving of
all 13 $B_u$ modes and 15 $A_g$ modes to dynamic instability of
investigated ferroelectrics. However, obviously important role also
belongs to the nonlinear interaction with participation of $A_u$ and $B_g$
nonsymmetric modes. For the point group $2/m$ in
addition to the $A_gB_u^2$ fully symmetric combination, the
invariants of $A_uB_gB_u$ type are also present. A significant role of such
nonlinear mixing of lattice vibrations is reflected in strong
internal deformations of \PS clusters at the phase transition. In
the crystal electron structure, such invariants obviously replicate
the hybridization of molecular orbitals of \PS clusters with
participation of atomic $d$ orbitals. Such hybridization is clearly
illustrated by the spatial distribution of electron density for the
energy level number 20, which is aligned from tin atoms to middle of
P~---~P bond (Fig.~\ref{fig8}).

The energy decreasing for the $s$ orbitals of phosphorus and sulfur at
the VB bottom (their contribution is dominated in the lowest
subbands~-- from I to V), and also lowering of energy for tin $s$
orbitals, with their contribution across whole range of the VB
(Fig.~\ref{fig6}), are obviously essential for energetic motivation
of the transition into the ferroelectric phase. Also some
lowering of the $p$ and $d$ orbitals energy occurs.

The electronic structure XPS measurements for $\rm Sn_2P_2S_6$
crystals~\cite{Grigas09,Grigas08} revealed the chemical shifts of Sn
and P electronic core states to a higher binding energy and of S
states to a lower binding energy at the crystal lattice formation.
This shift suggests a charge transfer from Sn and P to S atoms. The
binding energies and chemical shifts strongly change at the phase
transition. In the ferroelectric phase, the chemical shifts of Sn
and P atoms are higher while for S atoms they are smaller. So, for
all atoms of crystal structure at transition into the ferroelectric
phase the binding energy for core orbitals increases. These data
give evidence about localization of electron charge in space between
atoms, or about enhancement of chemical bonds covalency. Such
variation of the core orbitals energy agrees with experimentally
observed and founded at calculations (Fig.~\ref{fig4},\ref{fig6})
transformation of the VB structure and they support an energetic
stability of the ferroelectric phase.

Since covalency increases, there might be a possibility for the
drastic collapse of the sulfur ionic size (which is related to
charge transfer). For the $\rm S^{2-}$ ion the ionic radius is
1.84~\AA, the covalent radius $\approx 1.02$ \AA~\cite{webelements}.
If the size of sulfurs were small compared to the allowed space then
the sulfur atoms would be weakly bounded in the lattice. In this
case an imbalance between the decreased (due to small S radius)
repulsive forces and the polarization forces, what tend to displace
the ion from its position, also support structure rearrangement.

In whole, for \SPS crystal at transition from the paraelectric phase to
the ferroelectric one complex evolution of electron and phonon spectra
occurs, which could be presented as sequence of five steps.
Evidently as \emph{first} factor assists change of the electron
density charge distribution in the elementary cell by fully
symmetric breathing modes $A_g$. Such redistribution of the electron
density prompts the stereochemical activity of tin cations electron
lone pair and produces the covalence bonds of tin atoms with sulfur
atoms (hybridization of $sp^2d^5$ type), and also with phosphorus
atoms, what could be considered as \emph{second} part.

As \emph{third} stage could be considered the weakening of the
short-range repulsion between cations of tin and phosphorus, in
result of their charges lowering, and at significant coulomb
repulsion of the nearest (related by the inversion center) tin
cations. Mentioned the second and the third factors represent the
nonlinear interaction of $A_gB_u^2$ type, they govern an anisotropy
of polar shifting of atoms in the elementary cell and define
appearance of the dipole structure motives (---~Sn~---~$\rm
P_2S_6$~---), which are related to the polar normal coordinates of
$B_u$ symmetry. The \emph{fourth} important factor is the
dipole-dipole interaction which correlates orientation of local
dipoles (pseudospins) and defines appearance of the spontaneous
polarization in the crystal structure. And, at finish, as
\emph{fifth} circumstance must be accounted that all low symmetry
modes participate in result of permitted nonlinear $A_uB_gB_u$
relation in the structure transformation. Such combination of
structure deformations mirrors participation of phosphorus and
sulfur $d$ orbitals in the covalent bonds of \SPS crystal.

It is interesting to compare the peculiarities of chemical bonds in
\SPS sulfide and \SPSe selenide compounds, and also in the lead
contained $\rm Pb_2P_2S_6$ crystal. At first, we will consider the
binary compounds MX, where M~-- metal Ge, Sn, Pb, X~-- chalcogen O,
S, Se, Te. The stereoactivity of electron lone pair for metal atoms
is determined by $sp^2$ hybridization of their $s$ and $p$ orbitals
with $p$ orbitals of chalcogen atoms. Such hybridization is
determined by the positions of the energy levels of electron states
and by width of the related energy bands in the crystal
structure~\cite{Waghmare03,Walsh05}. The smallest energy difference
is present between positions of energy levels of germanium $s$
orbitals and oxygen $p$ orbitals. Consequently for the compound GeO,
the largest stereoactivity of $4s^2$ electron orbitals of Ge is
observed~\cite{Walsh05}. At transition from Ge to Sn and than to Pb,
the energy of chemical binding for the $s$ orbitals increases. Thus
with transition from O to S, and further to Se and Te, the energy of
chemical binding for their $p$ orbitals decreases. It is expected
that the hybridization of Ge $4s$ orbitals and O $2p$ orbitals is
the strongest, and hybridization for Pb $6s$ orbitals and Te $5p$
orbitals is the most weak. However the hybridization is also
influenced by width of related energy bands in the crystal
structure. The increase of width for the $s$ and $p$ electron states
energy bands and their overleap could partially compensate increase
of the energy distance between related energy levels, what produce
some level of the stereoactivity and covalency of M~---~X bonds.

Thus, at transition from \SPS to $\rm Pb_2P_2S_6$ the binding energy
for Pb $6s$ level increases, what make weaker the stereoactivity of
the $6s^2$ electron lone pair in the dodecahedron of sulfur atoms.
Obviously by this matter, the observed~\cite{Baltrunas95} ionicity for
the Pb~---~S bonds is higher in compare with Sn~---~S bonds
ionicity. The melting temperature and the energy gap both rise at
substitution Sn by Pb. The paraelectric phase in $\rm Pb_2P_2S_6$ is
stable at cooling till 4.2~K~\cite{Vysochanskii06}.

The largest stereoactivity of Ge $4s^2$ electron lone pair in surround
of sulfur atoms gives natural explanation of absence of $\rm
Ge_2P_2S_6$ crystal structure. The Ge atoms couldn't be placed in
positions with high coordination of sulfur atoms. Obviously
introducing into \SPS crystal impurity of germanium in the charge
state $\rm Ge^{2+}$ strongly elevates the temperature of ferroelectric
phase transition what was observed by dielectric
investigations~\cite{Maior00}. Certainly the impurity in charge
state $\rm Ge^{4+}$ will not be stereoactive and will not support
rise of temperature interval for the ferroelectric phase existence.

According to just described tendencies, at transition from \SPS to
\SPSe the stereoactivity of Sn $5s^2$ electron lone pair in the
dodecahedron of selenium will be smaller than in the case of sulfide
compound. However, for the ternary compounds the ion-covalence bonds
Sn~---~S and Sn~---~Se are modified depending on peculiarities of
P~---~S and P~---~Se bonds in $\rm P_2S(Se)_6$ anion clusters. On
the M\"{o}ssbauer spectroscopy data~\cite{Baltrunas95} at
substitution of sulfur by selenium, the isomer shift for $\rm
{}^{119}Sn$ nucleus decreases what directly show on higher covalency
of Sn~---~Se bonds. The NMR spectroscopy for $\rm
{}^{119}Sn$~\cite{Apperley93,Francisco94} shows increase of the
resonance frequency from $-781.3$~ppm in \SPS to $-608$~ppm in $\rm
Sn_2P_2Se_6$. The NMR spectral line for $\rm {}^{31}P$ decreases its
frequency from 92.12~ppm in \SPS~\cite{Apperley93} to 28.7~ppm in
\SPSe~\cite{Francisco94}. These data provide evidence about lowering
of the electron density in vicinity of tin nucleus and about their
rise at phosphorus nucleus at transition from sulfide to selenide
compound. The mentioned tendency could be explained by smaller
electronegativity of selenium. Obviously the bonds P~---~Se are less
polar what improve higher electron charge of phosphorus atoms in
anionic clusters. Such situation probably also support more
effective hybridization between cluster molecular orbitals and tin
atomic orbitals (mostly the bonding hybridization of tin orbitals
with molecular orbitals that are localized in middle of P~---~P
bond) and increases the stereoactivity of tin lone pair of electrons
in \SPSe crystal.

The growth of covalency and weakening of electrostatic interactions
determine lowering of the melting temperature for \SPSe crystal in
compare with sulfide analog, define decrease of the energy gap and
govern smaller temperature of the ferroelectric phase
transition~\cite{Vysochanskii06}.
\newline

\section{Conclusions}

The appearance of the spontaneous polarization in \SPS compound is
accompanied by the significant changes of electronic structure that
are observed in all subbands of this crystal VB. At the transition
from the paraelectric phase to the ferroelectric one, the
significant changes also occur for the phonon spectra in the whole
frequency range~-- for both external and internal vibrations of the
crystal lattice. The complicate evolution of the energy spectra
could be represented by following contributions. The fully
symmetrical ('breathing') $A_g$ modes change the space distribution
of the electron density in the elementary cell. This one initiates
the stereochemical activity of the tin cations electron lone pair
and support creation of their covalence chemical bonds with sulfur
atoms (the hybridization of $sp^2d^5$ type), and with phosphorus
atoms also. Thus in result of the ionic charges lowering, the
short-range repulsion between tin and phosphorus cations decreases,
however the coulomb repulsion between tin cations still remains
strong enough. These factors reflect the nonlinear interaction of
$A_gB_u^2$ type, they determine the anisotropy of polar deformations
in the elementary cell and induce appearance of the dipole structure
motives (---~Sn~---~$\rm(P_2S_6)$~---), which are related to the
polar normal coordinates of $B_u$ symmetry. The dipole-dipole
interaction correlates orientation of the local dipoles
(pseudospins) and governs the spontaneous polarization of the
crystal structure. In the structure rearrangement, all low symmetry
modes take part~-- the possibility of the nonlinear linking of
$A_uB_gB_u$ type is obvious. This interaction correlates with
involving of tin and phosphorus $d$ orbitals into creation of the
covalent bonds in \SPS crystal lattice. Small difference of the
paraelectric and ferroelectric phase's energies and activation
redistribution of the electron charge at the spontaneous
polarization appearance determine the presence of the three-well
local potential in the \SPS ferroelectrics.

\begin{acknowledgments}
Part of presented in this paper results of calculations was
conducted with support of computational cluster of Institute of
Condensed Matter Physics, Lviv. Authors would like to thank staff of
this facility (T. Bryk, T. Patsahan) for technical assistance. Also
authors are grateful to Dr. R. Yevych for helpful discussions.
\end{acknowledgments}

\let\textsc\WileyBibTextsc
\providecommand{\othercit}{} \providecommand{\jr}[1]{#1}
\providecommand{\etal}{~et~al.}

\end{document}